\documentclass[preprint2,twocolappendix]{aastex6}

\newcommand{\cs}{c_\mathrm{s}}

\newcommand{\pcc}{\mathrm{cm}^{-3}}
\newcommand{\kms}{\mathrm{km~s^{-1}}}
\newcommand{\pc}{\mathrm{pc}}
\newcommand{\myr}{\mathrm{Myr}}

\newcommand{\nave}{\langle n_0\rangle}

\newcommand{\rhoave}{\langle \rho_0\rangle}

\newcommand{\Bsh}{B_\mathrm{sh}}

\newcommand{\thetacr}{\theta_{\mathrm{cr}}}
\newcommand{\Thetacr}{\Theta_{\mathrm{cr}}}

\newcommand{\alfvenic}{Alfv{\'e}nic}
\newcommand{\alfven}{Alfv{\'e}n}

\newcommand{\nth}{n_\mathrm{th}}
\newcommand{\nmax}{n_\mathrm{max}}

\newcommand{\masscl}{M_\mathrm{cl}}
\newcommand{\Bcl}{B_\mathrm{cl}}

\newcommand{\Mcl}{M_\mathrm{cl}}
\newcommand{\ncl}{ {n}_\mathrm{cl}}
\newcommand{\rhocl}{\rho_\mathrm{cl}}
\newcommand{\Rcl}{R_\mathrm{cl}}
\newcommand{\Vcl}{V_\mathrm{cl}}

\newcommand{\ducl}{\delta u_\mathrm{cl,1D}}

\newcommand{\ngrav}{ n_\mathrm{grv}}

\newcommand{\Egrav}{ {\cal E}_\mathrm{grv}}
\newcommand{\Eth}{ {\cal E}_\mathrm{th}}

\newcommand{\Emag}{ {\cal E}_\mathrm{mag}}
\newcommand{\Ekin}{ {\cal E}_\mathrm{kin}}

\newcommand{\mucr}{\mu_\mathrm{cr}}

\newcommand{\alphath}{\alpha_\mathrm{th}}

\fullcollaborationName{The Friends of AASTeX Collaboration}

\begin{document}


\title{
Universal Properties of Dense Clumps in 
Magnetized Molecular Clouds Formed through Shock Compression of Two-phase Atomic Gases 
}


\author{
Kazunari iwasaki\altaffilmark{1}, 
and
Kengo Tomida\altaffilmark{2}
}


\altaffiltext{1}{
Center for Computational Astrophysics/Division of Science, 
National Astronomical Observatory of Japan, Mitaka, Tokyo 181-8588, Japan
}
\altaffiltext{2}{
Astronomical institute, Tohoku University, Aoba, Sendai 980-8578, Japan
}

\begin{abstract}
We investigate the formation of molecular clouds from atomic gas by using 
three-dimensional magnetohydrodynamical simulations, 
including non-equilibrium chemical reactions, heating/cooling processes, and self-gravity by 
changing the collision speed $V_0$ and the angle $\theta$ between the magnetic field and colliding flow.
We found that the efficiency of the dense gas formation depends on $\theta$.
For small $\theta$, anisotropic super-Alfv\'enic turbulence delays the formation of gravitationally unstable clumps.
An increase in $\theta$ develops shock-amplified magnetic fields along which the gas is accumulated,
making prominent filamentary structures.
We further investigate the statistical properties of dense clumps identified with different density 
thresholds. 
The statistical properties of the dense clumps with lower densities depend on $V_0$ and $\theta$ because
their properties are inherited from the global turbulence structure of the molecular clouds. 
By contrast, denser clumps appear to have asymptotic universal statistical properties, 
which do not depend on the properties of the colliding flow significantly. 
The internal velocity dispersions approach subsonic and plasma $\beta$ becomes order of unity.
We develop an analytic formula of the virial parameter 
which reproduces the simulation results reasonably well.
This property may be one of the reasons for the universality of the initial mass function of stars.

\end{abstract}


\section{Introduction} \label{sec:intro}

The formation and evolution of molecular clouds (MCs)
are crucial to set the initial condition of star formation.
The formation of MCs arises from accumulation of the diffuse atomic gas, 
which consists of the cold neutral medium (CNM) and warm neutral medium (WNM) \citep{Field1969}.

The formation of MCs has been intensively studied by numerical 
simulations of atomic colliding flows 
\citep[e.g.,][]{Heiles2005,Vaz2006,Henn2008,Inoue2012}.
The colliding flows model accumulation of the atomic gas owing to 
supernova explosions \citep[e.g.,][]{MO1977},
expansion of super-bubbles \citep[e.g.,][]{McCray1987},
galactic spiral waves \citep[e.g.,][]{Wada2011}, and 
large-scale colliding flows.
The shock compression induces the thermal instability \citep{Field1965} to form cold clouds 
\citep{Henn2000,Ki2000}.
At the same time, a part of the kinetic energy of the colliding flows is converted 
into the turbulent energy of the cold clouds and intercloud gas, generating 
supersonic translational velocity dispersions of the cold clouds \citep{KI2002}.
MCs form if a sufficient amount of the gas is accumulated into the cold clouds. 

Magnetic fields play a crucial role on the physical properties of MCs.
\citet{iwasaki2019} (hereafter Paper I) 
investigated the early stage of the molecular cloud formation through 
compression of the two-phase atomic gas with various compression conditions
\citep[also see][]{ii2009,Heitsch2009,Kortgen2015,DobbsWurster2021}.
Paper I found that the physical properties of the post-shock layers strongly depend on 
the angle $\theta$ between the magnetic field and collision velocity.
For a sufficiently small $\theta$, accumulation of the two-phase atomic gas drives 
super-Alfv\'enic turbulence \citep{Inoue2012}.
As $\theta$ increases, the Alfv\'en mach number of the turbulence decreases, and 
the shock-amplified magnetic field plays a major role in the dynamics of the post-shock layer.

In this paper, we investigate how the $\theta$-dependence of the turbulent properties 
affects the subsequent star formation 
by using the numerical simulations including self-gravity.
We focus on the physical properties of dense clumps that are often 
discussed by using the virial theorem both observationally \citep{BertoldiMcKee1992} and 
theoretically \citep{McKeeZweibel1992}.
In most colliding-flow simulations, 
the virial theorem of dense clumps have not been discussed quantitatively.
For instance, \citet{Inoue2012} investigated the physical properties of dense clumps and 
discuss their stability using the virial theorem, but 
their simulations did not include self-gravity and they did not provide quantitative discussions.
We will derive an analytic estimate of each term in the virial theorem on the basis of 
the results of the numerical simulations.

This paper is structured as follows:
In Section \ref{sec:method}, we describe the methods and models adopted in this paper.
We present our results including the global physical properties of the post-shock layers and 
the statistical properties of dense clumps in Section \ref{sec:results}.
Discussions are made in Section \ref{sec:discussion}.
Finally, we summarize our results in Section \ref{sec:summary}.

\section{Methods and Models}\label{sec:method}
\subsection{Methods}\label{sec:basic}

The basic equations are the same as those used in Paper I except that 
self-gravity is taken into account in this work. 
The basic equations are given by 
\begin{equation}
\frac{\partial \rho}{\partial t} + \frac{\partial \left( \rho v_i \right)  }{\partial x_i}= 0,
\label{eoc}
\end{equation}
\begin{equation}
\frac{\partial \rho v_i}{\partial t} 
+ \frac{\partial }{\partial x_j} \left( \rho v_i v_j + T_{ij} \right) = -\rho\frac{\partial \phi}{\partial x_i},
\end{equation}
\begin{equation}
\frac{\partial E}{\partial t} 
+ \frac{\partial }{\partial x_i} 
\left[ \left(E\delta_{ij} + T_{ij} \right) v_j  
- \kappa \frac{\partial T}{\partial x_i}\right] = - \rho v_i \frac{\partial \phi}{\partial x_i}-{\cal L},
\end{equation}
\begin{equation}
\frac{\partial B_i}{\partial t} 
+ \frac{\partial }{\partial x_j} \left(  v_j B_i - v_i B_j \right)=0,
\label{induc}
\end{equation}
\begin{equation}
    \nabla^2 \phi = 4\pi G \rho,
    \label{poisson}
\end{equation}
and
\begin{equation}
     T_{ij} = \left( P + \frac{B^2}{8\pi} \right)\delta_{ij}  - \frac{B_i B_j}{4\pi},
     \label{Tij}
\end{equation}
where $\rho$ is the gas density, $v_i$ is the gas velocity, $B_i$ is the magnetic field, 
$E=\rho v^2/2 + P/(\gamma-1) + B^2/8\pi$ is the total energy density, 
$\phi$ is the gravitational potential, $T$ is the gas temperature, 
$\kappa$ is the thermal conductivity,
and ${\cal L}$ is a net cooling rate per unit volume.
The thermal conductivity for neutral hydrogen is given by 
$\kappa(T) = 2.5\times 10^3 T^{1/2}$~cm$^{-1}$~K$^{-1}$~s$^{-1}$ \citep{Parker1953}.

The Possion equation (Equation (\ref{poisson}))
is solved by the multigrid method that has been implemented in Athena++ by 
\citet[][in preparation]{Tomida2022}. 
We have implemented the chemical reactions, radiative cooling/heating, ray-tracing of 
FUV photons into a public MHD code of {Athena++} \citep[][]{Stone2020} in Paper I.

We do not use the adaptive mesh refinement technique but use
a uniformly spaced grid.
The simulations are conducted with resolution of $2048\times 1024^2$, 
which is two times higher than that in Paper I.
The corresponding mesh size is $\Delta x\sim 0.02~$pc.
Although dense cores with $<0.1$~pc are not resolved sufficiently, 
the surrounding low-density regions, which is important for the MC formation 
and driving turbulence, are well resolved \citep{Kobayashi2020}.

To avoid gravitational collapse and numerical fragmentation, 
the cooling and heating processes are switched off 
if the local thermal Jeans length $\lambda_\mathrm{J}\equiv c_\mathrm{s}\sqrt{\pi/G\rho}$
is shorter than $4\Delta x$ \citep{Truelove1997}.
Since the temperature of the dense gas is about 20~K in our results,  
the Truelove condition sets the maximum density 
$n_\mathrm{max} \equiv 1.4\times 10^{5}~\mathrm{cm}^{-3}(N_x/2048)^2
(\cs/0.3~\mathrm{km~s^{-1}})^{-2}$ above which the gas is artificially 
treated as the adiabatic fluid with $\gamma=5/3$.

The maximum density $n_\mathrm{max}$ can be reached only when 
self-gravity becomes important.
As will be explained in Section \ref{sec:model}, we consider colliding flows of the 
two-phase atomic gas with collision speeds of $V_0=5~\kms$ and $V_0=10~\kms$. The cold phase of the atomic gas is as dense as 
$n_\mathrm{cold}\sim 50~\pcc$. 
The maximum densities obtained by the shock compression 
are estimated to be $n_\mathrm{cold}(V_0/0.3~\kms)^2\sim 10^4~\pcc$ for $V_0=5~\kms$ and 
$\sim 4\times 10^4~\pcc$ for $V_0=10~\kms$, both of which are smaller than $n_\mathrm{max}$.


Paper I took into account 9 chemical species 
(H, H$_2$, H$^+$, He, He$^+$, C, C$^+$, CO, $e$) following \citet{Inoue2012}.
We improved the chemical networks and thermal processes following \citet{Gong2017}.
We consider 15 chemical species 
(H, H$_2$, H$^+$, H$_2^+$, H$_3^+$, He, He$^+$, C, C$^+$, CH$_x$, CO, O, OH$_x$, HCO$^+$, $e$)
which are similar to those in \citet{Nelson1999}.
We calculate the far-ultraviolet (FUV) flux, which causes photo-chemistry and related thermal processes, 
is required to be calculated at any cells.
In the same way as in Paper I and \citet{Inoue2012}, FUV radiation is divided into
two rays which propagate along the $x$-direction from the left and right $x$-boundaries 
because the compression region has a sheet-like configuration, and 
the column densities along the $x$-direction are smaller than those in the $y$- and $z$-directions.
The two-ray approximation is also adapted 
for calculating the escape probabilities of the cooling photons.
The detailed description is found in Paper I.


The cosmic-ray ionization rate per hydrogen nucleus $\xi_\mathrm{H}$ varies significanly depending on regions. 
In this paper, we adopt the recently updated one in diffuse molecular clouds
$\xi_\mathrm{H} = 2\times 10^{-16}~$s$^{-1}$ \citep{Indriolo2007,Indriolo2015,NeufeldWolfire2017}, which 
is about ten times larger than that adopted in Paper I. 
We should note that the cosmic ray flux is attenuated toward the deep interior of molecular clouds, and
\citet{NeufeldWolfire2017} showed marginal evidence that 
$\xi_\mathrm{H}$ decreases with $A_\mathrm{V}$ in proportional to $A_\mathrm{V}^{-1}$ for $A_\mathrm{V}>0.5$. 
Our adopted value of $\xi_\mathrm{H}=2\times 10^{-16}~$s$^{-1}$ corresponds to an upper limit of 
the cosmic-ray ionization rate, and the gas temperature at dense regions may be slightly overestimated.

As a solver of the stiff chemical equations, we adopt the Rosenbrock method, which is 
a semi-implicit 4th-order Runge-Kutta scheme \citep{Press1986}.
The time width $\Delta t_\mathrm{chem}$ required to solve the chemical reactions with 
sufficiently small errors can be much smaller than that of the MHD part.
We integrate the chemical reactions with adaptive substepping, and
the number of the substeps is determined so that a error estimated from
3rd-order and 4th-order solutions becomes sufficiently small.


\subsection{A Brief Review of Paper I} \label{sec:review}

Before presenting the models in this study, we here review the results of 
Paper I that investigated the early stage
of molecular cloud formation from atomic gas by ignoring self-gravity.
They performed a survey in a parameter space of 
($\nave$, $V_0$, $B_0$, $\theta$),
where $\nave$ is the mean density of the pre-shock atomic gas, $V_0$ is the collision velocity,
and $B_0$ is the field strength, and $\theta$ is the angle between the collision velocity and 
magnetic field of the atomic gas.
At a given parameter set of $(\nave,B_0,V_0)$, they found that there is a critical angle 
$\thetacr$ which characterizes the physical properties of the post-shock layers.
They obtained an analytic formula of the critical angle as follows:
\begin{equation}
     \sin \thetacr = 0.1
     \left( \frac{\nave}{10~\mathrm{cm}^{-3}} \right)^{1/2}
     \left(  \frac{V_0}{5~\mathrm{km~s^{-1}}} \right)
     \left(  \frac{B_0}{3~\mu\mathrm{G}} \right)^{-1}.
     \label{th_crit}
\end{equation}

\begin{itemize}
\item For an angle smaller than $\thetacr$, 
 anisotropic super-Alfv\'enic turbulence is maintained by 
 the accretion of the two-phase atomic gas.
 The ram pressure dominates over the magnetic stress in the post-shock layers.
 As a result, a greatly-extended turbulence-dominated layer is generated.

\item For an angle comparable to $\thetacr$, the shock-amplified magnetic field weakens 
the post-shock turbulence, making the post-shock layer denser.
The mean post-shock density has a peak around $\theta\sim \thetacr$ as a function of $\theta$.

\item For an angle larger than $\thetacr$, 
      the magnetic pressure plays an important role in the post-shock layer.
      Their mean densities decrease with $\theta$ and 
      are determined by the balance between the post-shock magnetic pressure and 
      pre-shock ram pressure \citep[see also][]{Inoue2012}.

\end{itemize}

\subsection{Models}\label{sec:model}

We consider colliding flows of the two-phase atomic gas whose mean density is $\nave$ 
in the same manner as in Paper I.
The collision speed is $V_0$, magnetic field strength is $B_0$ and 
the angle between the colliding flow and magnetic field is $\theta$. 
The colliding flows move in the $\pm x$ directions, and the magnetic fields
are tilted toward the $y$-axis.

In order to generate the initial conditions for performing the colliding-flow simulations, 
the upstream two-phase atomic gas is generated using the same manner as in Paper I.
We prepare a thermally unstable gas in thermal equilibrium with 
a mean density of $\nave$ and a uniform magnetic field of $\mathbf{B}_0$
in a cubic simulation box of $-10~\pc \le x,y,z \le 10~\pc$ with the periodic boundary conditions 
in all the directions. 
As the initial condition, 
we add a velocity dispersion having a Kolmogrov spectrum of $0.64~\kms~(L/~\mathrm{pc})^{0.37}$, where 
$L$ is the size of regions.
The initial unstable gas evolves into the two thermally stable states, or the CNM and WNM, 
in about one cooling timescale. 
We terminate the simulations at $t=8~$Myr or $\sim 4$ cooling timescale of the initial unstable gas.
The density distribution of our upstream gas is highly inhomogeneous, where the dense CNM clumps are embedded 
in the diffuse WNM (see Figure 1 in Paper I).

In the colliding flow simulations, the simulation box is doubled in the $x$-direction, or 
$-20~\pc\le x\le 20\pc$, $-10~\pc \le y,z \le 10~\pc$.
In the $y$- and $z$-directions, periodic boundary conditions are imposed.
The $x$-boundary conditions at $x=\pm 20~$pc are imposed so that 
the initial distribution is periodically injected into the computation region from 
both the $x$-boundaries with a constant speed of $V_0$.

The model parameters are listed in Table \ref{tab:model}.
In all the models, $\nave=10~\pcc$ is adopted.
As in Paper I, 
we here consider the formation of MCs through accretion of HI clouds 
with a mean density of $\sim 10~\pcc$ \citep{Blitz2007,Fukui2009,Inoue2012,Fukui2017} rather than
through accretion of the warm neutral medium.
For all the models,
the field strength $B_0$ is set to $3~\mu$G, which 
is roughly consistent with a typical field strength in the atomic gas \citep{Heiles2005,Crutcher2010}.

As a fiducial model in this paper, we consider $V_0=5~\kms$.
In order to investigate the dependence of the results on the collision velocity, 
we perform additional simulations with $V_0=10~\kms$.
Our simulations model accumulation of the atomic gas by expansion of HI shells or
spiral shock waves \citep{McCray1987,Wada2011}.
The critical angle for the fiducial model is $\sin\thetacr=0.1$ for $V_0=5~\kms$ and 
$0.2$ for $V_0=10~\kms$ (Equation (\ref{th_crit})).

For each $V_0$, we consider cases with $\theta=0.25\thetacr$ and $\theta=2\thetacr$
as representative models of the turbulence-dominated and magnetic-pressure-dominated layers, respectively (Section \ref{sec:review}). 
We name a model by attaching ``$\Thetacr$'' in front of a value of $\theta/\thetacr$ followed by 
``V'' in front of a value of $V_0/1~\kms$ (Table \ref{tab:model}).

\begin{table}
    \begin{center}
    \begin{tabular}{|l|c|c|c|}
        \hline
        model  &  $V_0~[\kms]$ & $\theta$ & $t_\mathrm{first}$\\
        \hline
        \hline
        $\Thetacr 0.25 \mathrm{V5}$ & 5 & $1.5^\circ~(0.25 \thetacr)$ &  8.3~Myr\\
        \hline
        $\Thetacr 2\mathrm{V5}$ & 5 & $11^\circ~(2 \thetacr)$ & 6.0~Myr \\
        \hline
        $\Thetacr 0.25 \mathrm{V10}$ & 10 & $3^\circ~(0.25 \thetacr)$ & 8.8~Myr \\
        \hline
        $\Thetacr 2\mathrm{V10}$ & 10 & $24^\circ~(2 \thetacr)$ & 6.0~Myr \\
        \hline
     \end{tabular}
    \end{center}
    \caption{List of the model parameters. 
        In the rightmost column, $t_\mathrm{first}$ indicates the epoch when the first unstable clump is formed.
    In all the models, $\nave=10~\pcc$ and $B_0=3~\mu$G are adopted.}
    \label{tab:model}
\end{table}

It is useful to estimate global mass-to-flux ratios $\mu_\mathrm{layer}$ of the post-shock layers
\citep{Strittmatter1966,Mous1976,Nakano1978,Tomisaka1988,Tomisaka2014}.
The total mass of the post-shock layer increases with time, following $M_\mathrm{layer} = 2\rhoave V_0 L^2 t $,
where $L=20~\mathrm{pc}$ is the box size along the $y$- and $z-$axes.
The magnetic flux held in the post-shock layer depends on direction.
In the plane perpendicular to the $x$-axis ($y$-axis), the magnetic fluxes 
$\Phi_x$ ($\Phi_y$) is 
expressed as $B_0 L^2\cos \theta $ ($2B_0 V_0 L t \sin \theta$).
The global mass-to-flux ratio is estimated as 
$\mu_\mathrm{layer} = \min(M_\mathrm{layer}/\Phi_x, M_\mathrm{layer}/\Phi_y)$. 
It increases with time as $M_\mathrm{layer}/\Phi_x\propto t$, but 
does not exceed $M_\mathrm{layer}/\Phi_y = \rhoave L/(B_0\sin\theta)$.
The global gravitational instability of the post-shock layers occurs if 
$\mu_\mathrm{layer}$ exceeds a critical value of 
$\mu_\mathrm{cr}=(2\pi\sqrt{G})^{-1}$.
The ratio is given by 
\begin{eqnarray}
    \frac{\mu_\mathrm{layer}}{\mucr}
    &=& 4 \nave_{1}B_3^{-1} \nonumber \\
    &\times&  \min\left\{
        V_{5} t_{10} \left( \frac{\cos \theta}{1} \right)^{-1},
        L_{20}\left( \frac{\sin \theta}{0.2} \right)^{-1}
\right\},
\label{globalmu}
\end{eqnarray}
where $\nave_\mathrm{1}=\nave/10~\pcc$, $B_3=B_0/3~\mu\mathrm{G}$, $V_5=V_0/5~$km$~$s$^{-1}$,
$L_{20}=L/20~$pc, and $t_{10}=t/10~\mathrm{Myr}$.
The mass-to-flux ratios of the post-shock layers increase in proportion to time and exceed unity at $t=2.5V_5^{-1}~\mathrm{Myr}$.
They continue to increase at least until $t=10~$Myr for the $\theta=0.25\thetacr$ models while 
they reach the maximum values of $2$ at $t=5$~Myr for model $\Thetacr2$V5 and 
$4$ at $t=2.5~$Myr for model $\Thetacr2$V10.
The star formation will be suppressed for largely-inclined compressions with 
$\sin\theta>0.8\nave_{10} B_3 L_{20}$ because $M_\mathrm{layer}/\Phi_y<\mucr$ unless 
magnetic dissipation processes, such as ambipolar diffusion and/or turbulent reconnection work.



\section{Results} \label{sec:results}
In this section, we show the simulation results up to $t=t_\mathrm{first}+0.5$~Myr, 
where $t_\mathrm{first}$ is the formation time of the first unstable clump, which will be defined in Section \ref{sec:selfgravity}.
In our simulations, the resolution is not high enough to resolve dense cores ($> 10^5~\pcc$) and 
the thermal processes are switched-off in the regions where the Jeans condition is violated \citep{Truelove1997} 
instead of inserting sink particles. 
Therefore, we focus on the early evolution of the clumps rather than performing long-term simulations in this paper.

\begin{figure}[htpb]
  \centering
  \includegraphics[width=8.0cm]{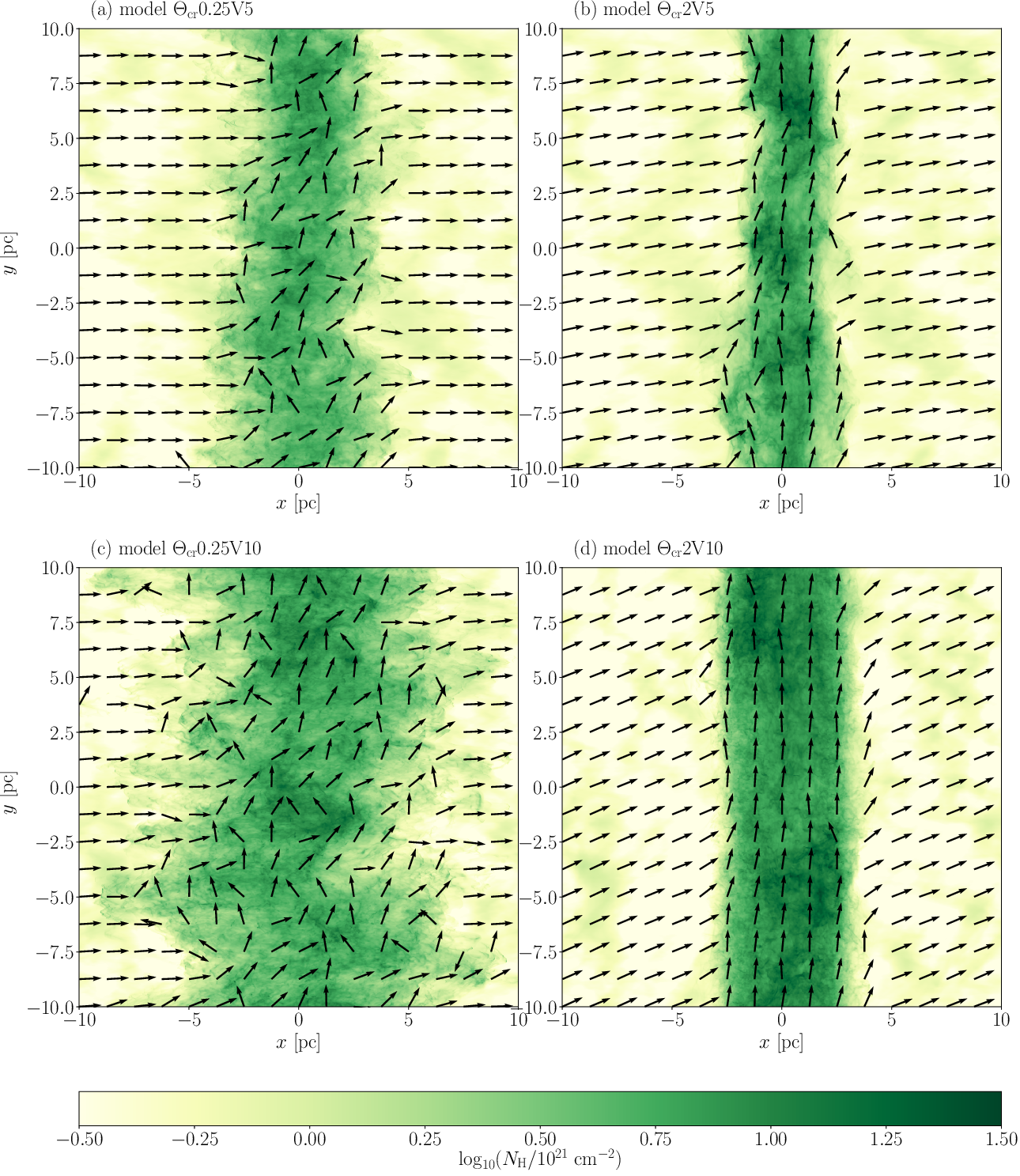}
\caption{
    Column density maps integrated along the $z$-axis 
    at $t=5~$Myr for models (a) $\Thetacr0.25\mathrm{V5}$, (b) $\Thetacr2\mathrm{V5}$, 
    (c) $\Thetacr0.25\mathrm{V10}$, and  
    (d) $\Thetacr2\mathrm{V10}$. 
    The black arrows indicate the direction of 
    the density-weighted mean magnetic field along the $z$-axis.
}
  \label{fig:col_5Myr}
\end{figure}

\subsection{Global Properties of post-shock Layers} \label{sec:globalprop}

\begin{figure}[htpb]
  \centering
 \includegraphics[width=9cm]{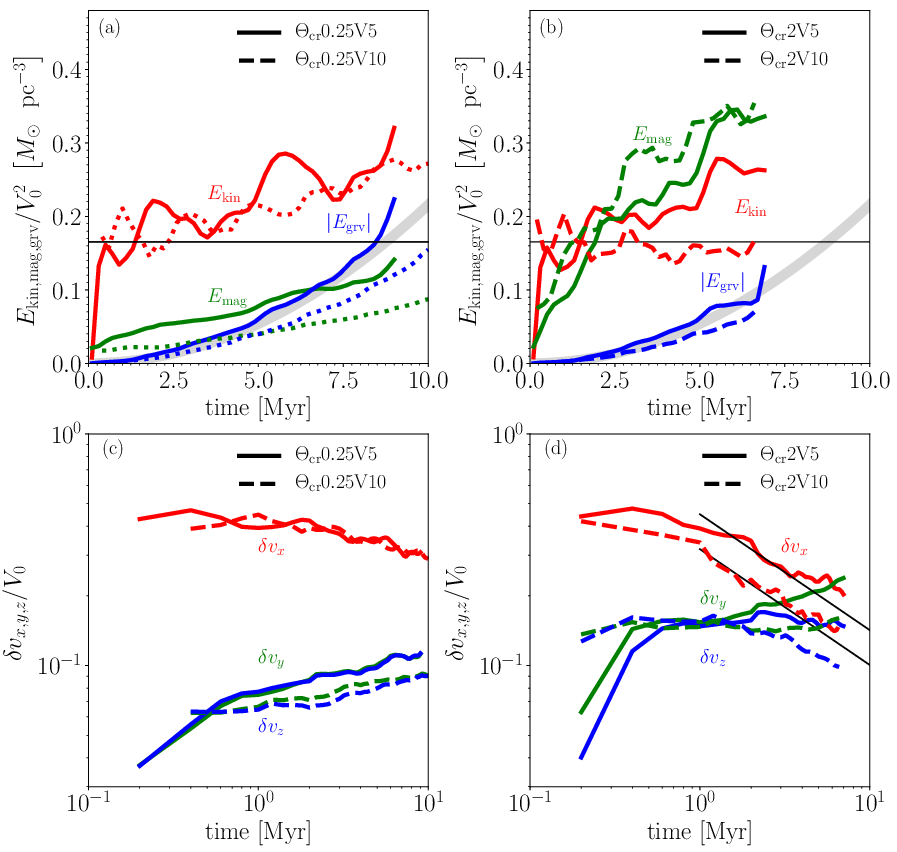}
\caption{
    (Top panels) Time evolutions of (red) kinetic energy, (green) magnetic energy, and (blue) 
    gravitational energy densities of the post-shock layers are
    compared between the models with (a) $\theta=0.25\thetacr$ and (b) $\theta=2\thetacr$.
    All the energy densities are divided by $V_0^2$.
    The gray line correspond to ${E}_\mathrm{grv,ana}$ shown in Equation (\ref{Egrav}).
    The horizontal black lines indicate the kinetic energy density flowing into the post-shock layer from the atomic gas, 
    $\rhoave V_0^2/2$.
    (Bottom panels) Time evolutions of the density-weighted velocity dispersions in the (red) $x$-, (green) $y$-, and 
    (blue) $z$-directions for (c) $\theta=0.25\thetacr$ and (d) $\theta=2\thetacr$.
    In Panel (d), the two black lines correspond to the predictions from $\delta v_x/V_0 = 0.45 \tau^{-1/2}$, where 
    $\tau=t\left( \nave/5~\pcc \right)\left( V_0/20~\kms \right)/(1~\mathrm{Myr})$ (Paper I) with $V_0=5~\kms$ and 
    $10~\kms$.
    In all the panels, the solid and dashed lines show the results with $V_0=5~\kms$ and $10~\kms$, respectively.
}
  \label{fig:eneevo}
\end{figure}

The early time evolution of the post-shock layers exhibits 
similar behaviors found in Paper I.
Figure \ref{fig:col_5Myr} shows the column densities integrated along the $z$-axis at $t=5$~Myr.
The accretion of the CNM clumps through the shock fronts disturbs the post-shock layers significantly. 
For $\theta=0.25\thetacr$, the CNM clumps are not decelerated 
by the Lorentz force significantly after passing through the shock fronts because the 
compression is almost parallel to the magnetic field (Figures \ref{fig:col_5Myr}a and \ref{fig:col_5Myr}c).
Figure \ref{fig:eneevo}a shows the time evolution of the mean (volume-averaged) kinetic, magnetic, and 
gravitational energy densities, ${E}_\mathrm{kin}$, ${E}_\mathrm{mag}$, ${E}_\mathrm{grv}$ of 
the post-shock layer\footnote{
In a similar way as Paper I, 
the post-shock layer is defined as the region confined by the two shock fronts
whose positions correspond to the minimum and maximum of the $x$ coordinates 
where $P(x,y,z)\ge P_\mathrm{th}$ is satisfied, where $P_\mathrm{th}$ is 
a threshold pressure. 
In this paper, we adopt a value of $P_\mathrm{th}=5.8\times 10^{3}~$K~cm$^{-3}$.} for $\theta=0.25\thetacr$. 
We first focus on the results with $V_0=5~\kms$.
The kinetic energies are always larger than the magnetic energies, 
indicating that the magnetic field lines are easily bent by the gas motion. 
This can be seen in the magnetic field directions that are random rather than 
organized as shown in Figures \ref{fig:col_5Myr}a and \ref{fig:col_5Myr}c.


An increase in $V_0$ from $5~\kms$ to $10~\kms$ does not change 
the physical properties of the post-shock layers for $\theta=0.25\thetacr$.
Figure \ref{fig:eneevo}a clearly shows that 
all the energy densities are roughly proportional to $V_0^2$, keeping the relative ratios constant.
This can be understood by the fact that the dynamics in the post-shock layers are driven by 
the ram pressure that is proportional to $V_0^2$.

Figure \ref{fig:eneevo}c shows that 
$\delta v_{x,y,z}/V_0$ does not depend on $V_{0}$ significantly, indicating that 
the velocity dispersions in the post-shock layers are proportional to the collision speed.
For both the models, the degree of anisotropy of the velocity dispersion, which is defined as $\delta v_x/\sqrt{(\delta v_y^2 + 
\delta v_z^2)/2}$, is as large as $\sim 3$ for $V_0=5~\kms$ and $\sim 4$ for $V_0=10~\kms$ at $t=5$~Myr.
In addition, $\delta v_z$ is almost identical to $\delta v_y$ for each model. 
This suggests that there is no preferential directions in the $(y,z)$-plane.



When $\theta$ is increased from $0.25\thetacr$ to $2\thetacr$ for $V_0=5~\kms$ (model $\Thetacr2\mathrm{V5}$), 
the physical properties of the post-shock layer drastically changes.
Comparison between Figures \ref{fig:col_5Myr}a and \ref{fig:col_5Myr}b
shows that the post-shock layer is thinner and denser 
for $\theta=2\thetacr$ because 
the Lorentz force owing to the shock-amplified magnetic field decelerates 
the CNM clumps efficiently for $\theta=2\thetacr$.
The shock compression amplifies the magnetic energy that  
dominates over the kinetic energy at $t>2~$Myr (Figure \ref{fig:eneevo}b).  
Figure \ref{fig:col_5Myr}b shows that 
the magnetic field directions in the post-shock layer are parallel to the $y$-axis.
At the early phase ($t<0.5~$Myr), Figures \ref{fig:eneevo}c and \ref{fig:eneevo}d show that 
$\delta v_x$ for model $\Thetacr2\mathrm{V5}$ is as large as that for 
model $\Thetacr0.25\mathrm{V5}$ because the magnetic field has not been amplified yet.
In the subsequent evolution, $\delta v_x$ decreases in proportion to $t^{-1/2}$ (Paper I), and 
all the three components $\delta v_{x,y,z}$ become 
of similar magnitudes around $t\sim 5$~Myr.

Unlike model $\Thetacr0.25\mathrm{V5}$, there is anisotropy in the velocity dispersion in the 
$(y,z)$ plane for model $\Thetacr2\mathrm{V5}$. The velocity dispersion along the $y$-axis
increases with time while $\delta v_z$ is constant with time.
The continuous increase in $\delta v_y$ 
corresponds to the formation of filamentary structure by self-gravity, 
which will be shown in Figure \ref{fig:colend}.

Unlike the models with $\theta=0.25\thetacr$,
the time evolution of the physical quantities depends on the collision speed for the $\theta=2\thetacr$ models. 
The relative importance of $E_\mathrm{mag}$ to the total energy is more significant for $V_0=10~\kms$ than for $V_0=5~\kms$.
After the early evolution ($t<1~\myr$) in which 
the velocity dispersions $\delta v_{x,y,z}$ increase in proportion to $V_0$,
$\delta v_x$ begins to decrease earlier for $V_0=10~\kms$ than for $V_0=5~\kms$.
This behavior is consistent with the results of Paper I where they found that 
$\delta v_x/V_0 = 0.3 \tau^{-1/2}$, where 
$\tau=t\left( \nave/5~\pcc \right)
\left( V_0/20~\kms \right)/(1~\mathrm{Myr})$
\footnote{
The results are better reproduced by multiplying $\delta v_x/V_0 = 0.3\tau^{-1/2}$ by 1.5.
This difference comes from the fact that it originally has a relatively large dispersion
as shown in the bottom panel of Figure 17 in Paper I.
The difference of cooling/heating processes between this paper and Paper I can be another reason.
}
The formula suggests that $\delta v_x/V_0$ is independent of $\rhoave$ and $V_0$ if $\delta v_x$
is measured at the time of reaching the same column density.
The effect of the rapid decrease in $\delta v_x$ can be seen in the fact that the shock fronts 
are flatter for $V_{0}=10~\kms$ (Figure \ref{fig:col_5Myr}b and \ref{fig:col_5Myr}d). 
The anisotropy of the velocity dispersions in the ($y,z$) plane exists also for the $V_0=10~\kms$.
Interestingly, the epochs ($t\sim 5~\myr$) when $\delta v_x$ becomes equal to $\delta v_y$ 
appear to be independent of $V_0$.



There are two main mechanisms to drive the post-shock turbulence.
One is the accretion of the upstream CNN clumps as mentioned above.
Since the momentum flux of the CNM clumps is much larger than 
the averaged post-shock momentum flux $\nave V_0^2$, 
the accretion of the CNM clumps significantly disturbs the post-shock layers (Paper I).
The other arises from deformation of the shock fronts that are generated by the interaction between the shock fronts 
and inhomogeneous upstream gas \citep{Kobayashi2020}.
The post-shock velocity can be super-sonic if the shock normal is tilted more than $\sim 30$ degrees \citep{LandauLifshitz1959}.

The effect of the thermal instability on the post-shock turbulence is minor.
\citet{Kobayashi2020} investigated how the post-shock turbulence depends on the amplitude of initial density 
perturbations.
They found that the thermal instability contributes to drive turbulence only when $\delta n_0/\nave$ is less than 
10\%, where $\delta n_0$ is the upstream density perturbation. 
In such situations, the energy conversion efficiency $\epsilon$ from the upstream kinetic energy $\rhoave V_0^2/2$ 
to the post-shock turbulence energy is as low as a few percents.
When $\delta n_0/n_0>10~\%$, the interaction between shock fronts and density inhomogeneity drives 
strong turbulence where $\epsilon$ is as high as 10\%, which 
is roughly consistent with the values obtained in this paper, which is 
$\epsilon \sim (\delta v_x/V_0)^2 \sim 16\%$ (Paper I).


\subsection{Structure Formation due to Self-gravity}\label{sec:selfgravity}

%
%
%
%

Figures \ref{fig:eneevo}a and \ref{fig:eneevo}b show the 
time evolution of the gravitational energy densities of the post-shock layers that are estimated by 
\begin{equation}
    E_\mathrm{grv}= -\left( \int_\mathrm{layer} 
    d^3x \right)^{-1} L^2 \int_\mathrm{layer} \rho x \frac{\partial \phi}{\partial x} dx,
\end{equation}
where the integration is done over the post-shock layer\footnote{
    We here estimate the gravitational energy due to the $x$-component of the 
    self-gravitational force only because 
the periodic boundary conditions are imposed in $y$ and $z$.
}.
The continuous accumulation of the atomic gas deepens the gravitational potential well, 
and $-E_\mathrm{grv}$ increases with time.
The total energies of the post-shock layers however are positive, indicating that 
the post-shock layers remain unbound until at least the end of the simulations.

Assuming that the gas density is uniform inside the post-shock layer, 
one can derive the gravitational energy density analytically as follows:
\begin{equation}
    {E}_\mathrm{grv,ana} = -\frac{\pi G \Sigma_0(t)^2}{3},
    \label{Egrav}
\end{equation}
where $\Sigma_0(t) = 2\rhoave V_0 t$ is the mean column density of the post-shock layer.
Figures \ref{fig:eneevo}a and \ref{fig:eneevo}b show that 
the time evolution of ${E}_\mathrm{grv}$ is consistent with 
that of ${E}_\mathrm{grv,ana}$ for all the models, indicating that the gravitational energy densities 
increase in proportion to $V_0^2$.
The deviations of $E_\mathrm{grv}$ from $E_\mathrm{grv,ana}$ come from 
the non-uniform density distributions.

\begin{figure}[htpb]
  \centering
 \includegraphics[width=8cm]{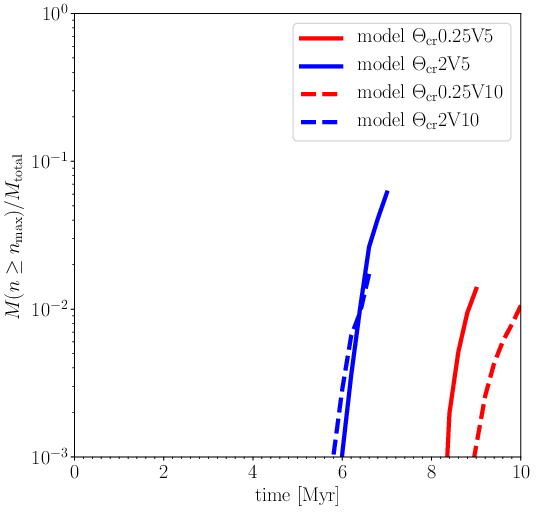}
\caption{
Time evolution of the mass fraction of dense gas with 
$n>n_\mathrm{max}$ for models (solid red) $\Thetacr0.25$V5, (solid blue) $\Thetacr2$V5, 
(dashed red) $\Thetacr0.25$V10, and (dashed blue) $\Thetacr2$V10.
}
  \label{fig:mass_nth}
\end{figure}

As an indicator of star formation, we measure the mass fraction of the dense gas whose 
density is larger than $n_\mathrm{max}$ above which the cooling/heating processes are switched off to 
prevent gravitational collapse (Section \ref{sec:model}).

The time evolution of the mass fraction of the dense gas
with $n>n_\mathrm{max}$ does not depend on $V_0$ significantly for each $\theta$ (Figure \ref{fig:mass_nth}). 
This behavior can be understood by the fact that the ratios of the thermal, kinetic, and magnetic energies 
to the gravitational energy are independent of $V_0$ (Figure \ref{fig:eneevo}).
We expect that an increase of $\nave$ accelerates the formation of the dense gas with $n>\nmax$ because 
the gravitational energy density is proportional to $\nave^2$ while the ram pressure of the pre-shock gas 
is proportional to $\nave$.


As is expected in Paper I, the formation of the dense gas with $n>n_\mathrm{max}$ is delayed by compressions with smaller $\theta$
because anisotropic super-Alf\'enic turbulence is driven for $\theta$ less than $\thetacr$.

Another interesting feature is that the difference in $V_0$ affects the formation of the dense gas with $n>n_\mathrm{max}$ 
in a different way at different $\theta$. As $V_0$ increases, the time evolution of the dense gas fraction 
does not change significantly for $\theta=2\thetacr$ while 
the formation of the dense gas is suppressed slightly for $\theta=0.25\thetacr$.
This is due to the fact that the $\theta$-dependence of the velocity field differs depending on the collision velocity.
For $\theta=0.25\thetacr$, the anisotropy of the velocity dispersion becomes more significant 
with increasing $V_0$ (Figure \ref{fig:eneevo}c).



We define $t_\mathrm{first}$, the time when the first bound clump is formed, as the earliest time at which 
the total energy of a clump with $n>n_\mathrm{max}$ becomes negative,
where the clumps are identified by a friends-of-friends method.
Table \ref{tab:model} lists $t_\mathrm{first}$, which are almost identical to the epochs when 
the mass with $n>n_\mathrm{max}$ increases rapidly as shown in Figure \ref{fig:mass_nth}.

\begin{figure}[htpb]
  \centering
  \includegraphics[width=8.5cm]{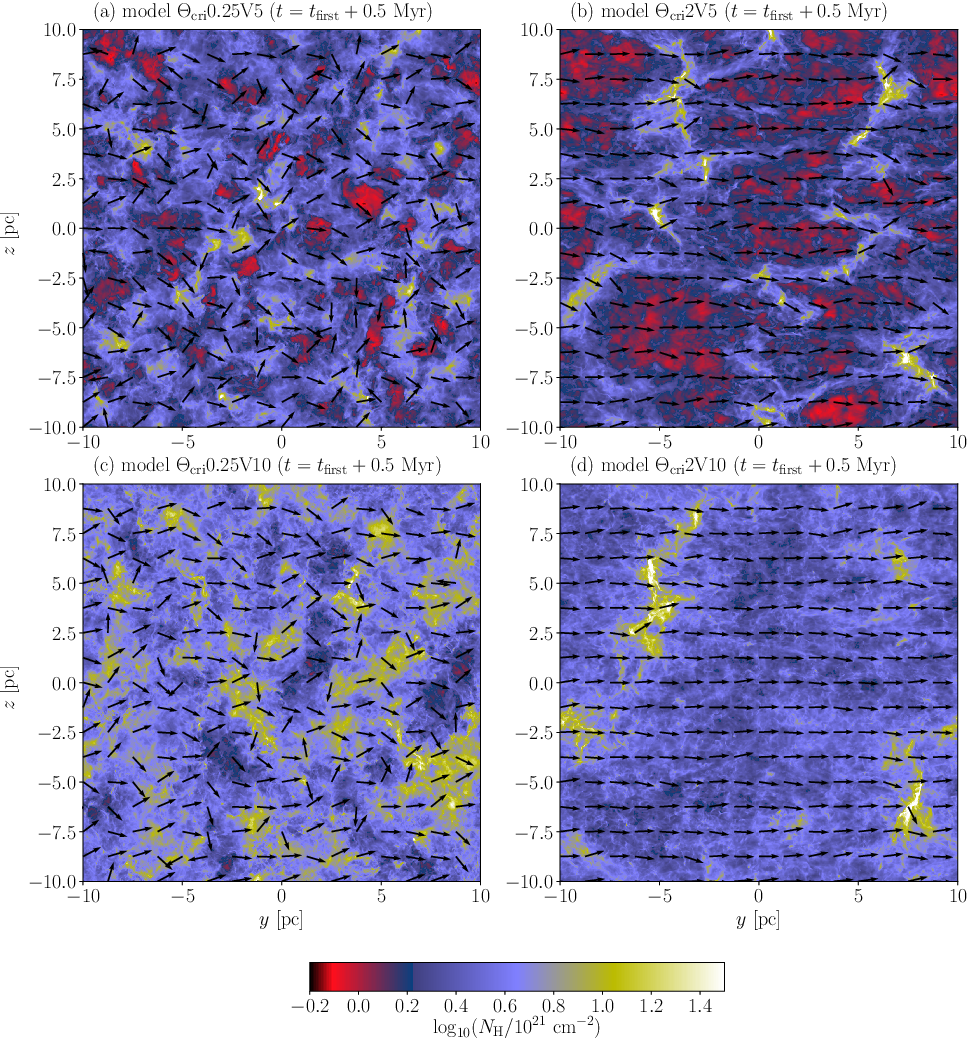}
\caption{
    Color maps of the column densities integrated along the $x$-axis for 
    models (a)$\mathrm{\Thetacr0.25V5}$, (b)$\mathrm{\Thetacr2V5}$, (c)$\mathrm{\Thetacr0.25V10}$, and 
    (d)$\mathrm{\Thetacr2V10}$ at $t_\mathrm{first}+0.5~\mathrm{Myr}$.
    In each panel, the black arrows indicate the direction of the magnetic fields.
}
  \label{fig:colend}
\end{figure}

The column density distributions along the collision direction  strongly depend on $\theta$ 
as shown in Figure \ref{fig:colend}.
For the models with $\theta=2\thetacr$, prominent filamentary structures develop, and  
their elongations are roughly perpendicular to the mean magnetic field 
(Figures \ref{fig:colend}b and \ref{fig:colend}d).
This is because the gas is accumulated by the self-gravity preferentially along with the magnetic fields.
By contrast, the column density maps for the models with $\theta=0.25\thetacr$ do not show filamentary structures clearly.
The projected magnetic fields are randomized.
The detailed structure of filamentary clouds is investigated in forthcoming papers.

%
%
%

\subsection{Field strength-density Relation}

\begin{figure}[htpb]
  \centering
 \includegraphics[width=7cm]{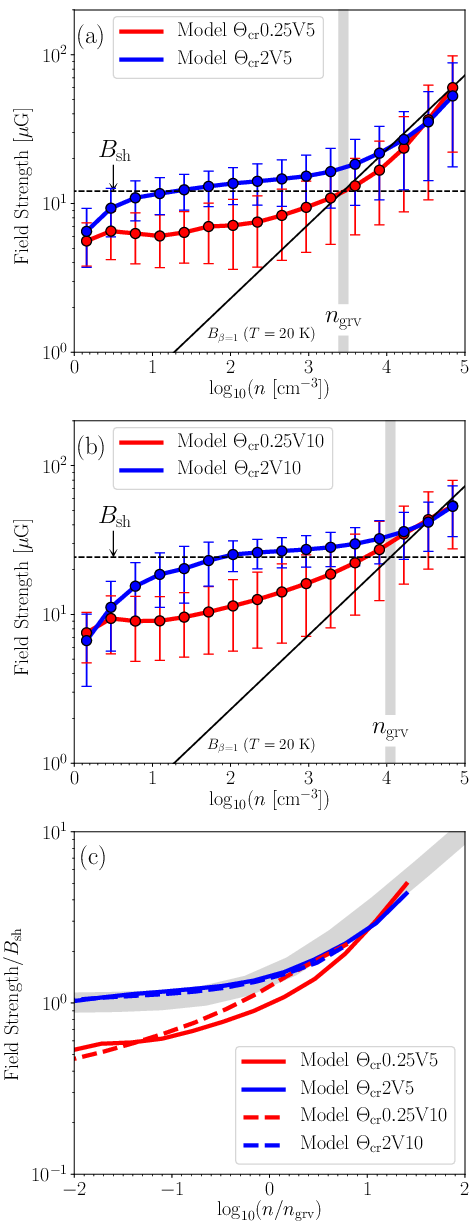}
\caption{
    Mean field strength at each number density bin for the models 
    with (a)$V_0=5~\kms$ and (b)$V_0=10~\kms$ at $t=t_\mathrm{first}+0.5$~Myr.
    Panels (a) and (b) compare the results with $\theta=0.25\thetacr$ and $2\thetacr$.
    The error bars correspond to the dispersion at a given density bin.
    The horizontal dashed lines indicate $B=\Bsh$.
    The field strength $B_{\beta=1}= \sqrt{8\pi \rho (0.3~\kms)^2}$, which is proportional to $n^{1/2}$,
    is plotted in the black solid line.
    Panel (c) is the same as Panels (a) and (b) but the horizontal and vertical axes are 
    normalized by $n_\mathrm{grv}$ and $B_\mathrm{sh}$, respectively.
    The thick gray line indicates the analytic formula shown in Equation (\ref{Bana}).
}
  \label{fig:Bn}
\end{figure}

The turbulence structures in the post-shock layers 
are reflected in the field strength-density ($B$-$n$) relation.
Figures \ref{fig:Bn}a and \ref{fig:Bn}b show 
the mean field strengths as a function of density 
at $t_\mathrm{first}+0.5~\mathrm{Myr}$ 
for the models with $V_{0}=5~\kms$ and $V_0=10~\kms$, respectively.

We first consider the models with $V_0=5~\kms$ (Figure \ref{fig:Bn}a).
In low densities less than $10^3~\pcc$,
model $\Thetacr2$V5 shows that the mean field strength is determined by 
the shock compression that amplifies the field strength up to
\begin{eqnarray}
 B_\mathrm{sh} &=& 
 \sqrt{8\pi \rhoave}V_0 \nonumber \\
 &=& 12~\mu\mathrm{G}~\left( \frac{\nave}{10~\pcc} \right)^{1/2} \left( \frac{V_0}{5~\mathrm{km~s^{-1}}} \right),
 \label{Bsh}
\end{eqnarray}
which is derived by balance between the pre-shock ram pressure 
and post-shock magnetic pressure.
The independence of $B\sim \Bsh$ on $n$ for model $\Thetacr2$V5 
reflects gas condensation along the magnetic field, corresponding to the formation of filaments.
By contrast, for model $\Thetacr0.25$V5, 
the mean field strength ($\sim 6~\mu$G) is amplified by shear motion owing to the anisotropic super-Alfv\'enic turbulence.
Assuming that the energy transfer efficiency from the kinetic energy to the magnetic energy $\epsilon$ is 20\%, 
the following estimated field strength $B_\mathrm{tur}$ is consistent with the simulation results,
\begin{equation}
    B_\mathrm{tur} = 5.4~\mu\mathrm{G}\left( \frac{\epsilon}{0.2} \right)^{1/2}\left( \frac{\nave}{10~\pcc} \right)^{1/2} \left( \frac{V_0}{5~\mathrm{km~s^{-1}}} \right).
    \label{Btur}
\end{equation}

Once the gas density becomes larger than $\sim 10^3~\pcc$, 
the $\theta$-dependence of the mean field strengths disappears, and the mean field strengths 
begin to increase with the gas density, following a similar $B$-$n$ relation.
Interestingly, the mean field strengths in the high density range 
are consistent with the field strength $B_{\beta=1}$,
\begin{equation}
    B_{\beta=1} = 23~\mu\mathrm{G}~\left( \frac{\cs}{0.3~\kms} \right) \left( \frac{n}{10^4~\pcc} \right)^{1/2},
\end{equation}
which is given by $\beta=1$ at $T=20~$K that is a typical temperature of the dense regions, where 
$\beta$ is the plasma beta.
This is where the self-gravity in the clumps becomes significant and the field amplification mode changes.

The $B$-$n$ relations of the models with $V_0=10~\kms$ behave similarly to those of 
the models with $V_0=5~\kms$.
At each $\theta/\thetacr$, the field strength increases by a factor of two
owing to the increase in $V_0$.
This is consistent with the predictions from Equations (\ref{Bsh}) and (\ref{Btur}).
Similar to the models with $V_0=5~\kms$, in the high density regions, 
the field strengths follow $B_{\beta=1}$ although the density range is limited. 
The critical density where the transition between the low and high density regimes occurs
is shifted toward high densities when $V_0$ increases from $5~\kms$ to $10~\kms$.

For the models with $\theta=2\thetacr$, the mean field strengths take a constant value of $\Bsh$ 
for lower densities and 
it increases following $B_{\beta=1}\propto n^{1/2}$ for higher densities (Figure \ref{fig:Bn}b).
From these facts, a critical density $\ngrav$ can be derived by equating 
$\Bsh$ and $B_{\beta=1}$ as follows:
\begin{eqnarray}
    \ngrav &=& \nave \left( V_0/\cs \right)^2 \nonumber \\
    &=& 
    2.8\times 10^3~\pcc~
    \left( \frac{\nave}{10~\pcc} \right)
    \nonumber \\
    && 
    \times \left( \frac{V_0}{5~\kms} \right)^2
    \left( \frac{\cs}{0.3~\kms} \right)^{-2}.
    \label{ngrav}
\end{eqnarray}
When the density exceeds $\ngrav$, the self-gravitational force amplifies the magnetic fields.

Figure \ref{fig:Bn}c is the same as Figures \ref{fig:Bn}a and \ref{fig:Bn}b but 
for the gas density and field strength are normalized by $\ngrav$ and $\Bsh$, respectively.
The $B$-$n$ relation is characterized by $\ngrav$ and $\Bsh$ reasonably well for the models with $\theta=2\thetacr$.
The mean field strength for the models is well approximated by 
\begin{equation}
    B_\mathrm{ana} = \Bsh \sqrt{ 1 + \beta_\mathrm{d}^{-1}(n/\ngrav)},
    \label{Bana}
\end{equation}
where $\beta_\mathrm{d}$ is a typical plasma $\beta$ for higher densities, and 
set to $1$ in Figure \ref{fig:Bn}.
Equation (\ref{Bana}) is similar to that found by \citet{Tomisaka1988} who 
investigate the magnetohydrostatic structure of axisymmetric clouds.

\section{Statistical Properties of Dense Clumps} \label{sec:clump}
In this section, we investigate the physical properties of dense clumps.
The clumps are identified by connecting adjacent cells whose densities are larger 
than a threshold number density of $\nth$.
We extract dense clumps consisting of more than 100 cells, whose internal structure 
is sufficiently well resolved at least with 6 cells along the diameter.
\citet{Dib2007} estimated errors for the terms in the virial theorem by using 
a uniform sphere, and found that the errors are not larger than $\sim 15~$\% if 
the diameter of the sphere is resolved by four cells.
\citet{Kobayashi2022} also found that the internal velocity dispersions of dense clumps 
are numerically converged if the clump size is resolved at least by 5 cells.
This is because the largest scale of a clump, which corresponds to the clump size, 
has the largest power of the internal turbulence spectrum.

\begin{figure}[htpb]
  \centering
 \includegraphics[width=9cm]{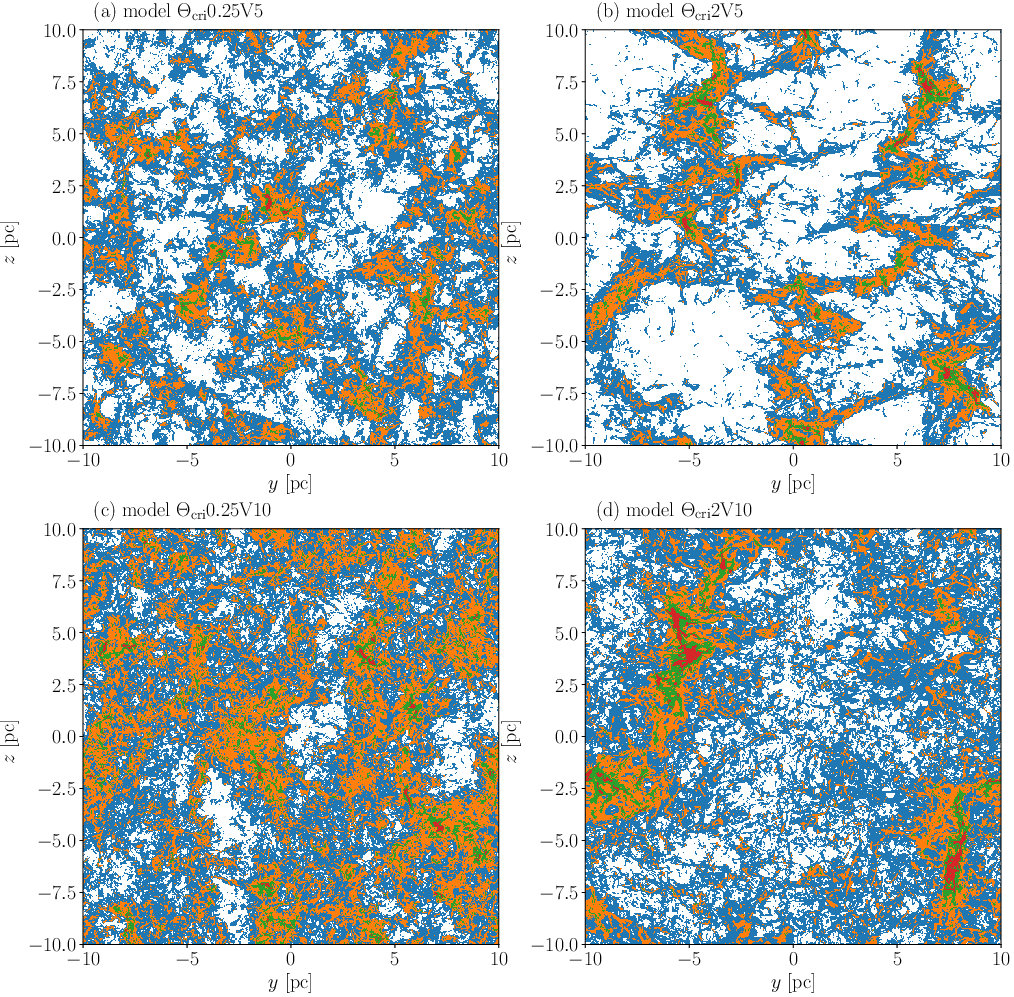}
\caption{
Spatial distributions of the dense clumps identified with four different threshold densities, 
$\nth=$ (blue)$10^3~\pcc$, (orange)$10^{3.5}~\pcc$, (green)$10^{4}~\pcc$, and (red)$10^{4.5}~\pcc$ 
for models (a)$\Thetacr$0.25V5, (b)$\Thetacr$2V5, (c)$\Thetacr0.25$V10, and (d)$\Thetacr$2V10 
at $t=t_\mathrm{first}+0.5~\mathrm{Myr}$.
}
  \label{fig:nth}
\end{figure}

We choose a minimum threshold density of $10^3~\pcc$ because 
we could not identify dense clumps as an isolated structure at lower densities.
We here consider the dense clumps identified with four different threshold densities
($\nth=10^3~\pcc$, $10^{3.5}~\pcc$, $10^4~\pcc$, and $10^{4.5}~\pcc$).
The spatial distributions of the dense clumps identified at $t=t_\mathrm{first}+0.5~\mathrm{Myr}$ 
are displayed in Figure \ref{fig:nth}.
The filamentary structures seen in the left panels of Figures \ref{fig:colend} correspond to 
the dense clumps with $\nth\ge 10^{3.5}~\pcc$.

We define the mass $M_\mathrm{cl}$, the position of the center of mass $\mathbf{x}_\mathrm{cl}$, and 
density-weighted mean velocity $\mathbf{v}_\mathrm{cl}$ of an identified clump as follows:
\begin{equation}
    M_\mathrm{cl} = \int_V \rho d^3x,
\end{equation}
\begin{equation}
    \mathbf{\bf x}_\mathrm{cl}= \frac{1}{\Mcl}\int_V \rho \mathbf{x} d^3x, 
\end{equation}
\begin{equation}
    \mathbf{\bf v}_\mathrm{cl}= \frac{1}{\Mcl}\int_V \rho \mathbf{v} d^3x,
\end{equation}
where $V$ denotes the volume of each identified clump.

In this paper, when considering the virial theorem, the surface terms are neglected although 
they can be important to judge the stability of dense clumps \citep{Dib2007}.
In a future paper, we will investigate the stability of dense clumps by taking into account the surface terms as well 
as the time evolution of filamentary structures.

The virial theorem without the surface terms is 
\begin{equation}
    \frac{1}{2}  \frac{d^2I}{dt^2} 
    = 2\left( \Eth + \Ekin \right) + \Emag + \Egrav 
    \label{virialE}
\end{equation}
where $I$ is the moment of inertia,
$\Eth$ and $\Ekin$ are the thermal and kinetic energies 
\begin{equation}
    \Eth = \frac{3}{2}\int_{V} P d^3x,\;\;\mathrm{and}\;\;
    \Ekin = \frac{1}{2}\int_{V} \rho \mathbf{u}^2 d^3x,
\end{equation}
and $\Emag$ is the magnetic energy,
\begin{equation}
    \Emag = \int_{V} \frac{B^2}{8\pi} d^3x,
\end{equation}
where $\mathbf{u} = (\mathbf{v}  - \mathbf{v}_\mathrm{cl})$.
In Equation (\ref{virialE}), 
the gravitational energy is denoted by 
\begin{equation}
    \Egrav = -\int_{V} \rho \mathbf{r} 
    \cdot \mathbf{\nabla} \phi d^3x,
\end{equation}
where $\mathbf{r} = (\mathbf{x}  - \mathbf{x}_\mathrm{cl})$.
$\Egrav$ is not exactly the same as the gravitational energy of the clump, 
and includes the tidal effect from the external gas.
In this paper, $\Egrav$ is called the gravitational energy.
It can be positive when the tidal force exerted by the external gas destroys the clump.

In preparation for interpreting the results, 
we express the energies in terms of the mean values,
\begin{equation}
    \Eth = \frac{3}{2} \Mcl c_\mathrm{s}^2,
    \label{Ethcl0}
\end{equation}
\begin{equation}
    \Ekin = \frac{3}{2} \Mcl \ducl^2,
    \label{Ekincl0}
\end{equation}
\begin{equation}
    \Emag = \frac{\Bcl^2}{8\pi} \frac{\Mcl}{\rhocl},
    \label{Emagcl0}
\end{equation}
\begin{equation}
    \Egrav \sim -\frac{3}{5}\frac{G\Mcl^2}{R_\mathrm{cl}} 
    \label{Egrvcl0}
\end{equation}
where $\cs$ is the mean sound speed, $\ducl$ is the one-dimensional internal 
velocity dispersion, $\Bcl$ is the mean field strength, 
$\Vcl$ is the volume of dense clumps, $\rhocl = \Mcl/\Vcl$ is the mean density.
In derivation of Equation (\ref{Egrvcl0}), we assume a uniform spherical cloud with a radius of $\Rcl$.
In this paper, as a size of clumps, we use
\begin{equation}
    \Rcl = \left( \frac{3\Vcl}{4\pi \rhocl} \right)^{1/3}.
\end{equation}

All the statistical quantities shown in this section are estimated using
the dense clumps identified in $t_\mathrm{first}-0.5~\mathrm{Myr}\le t\le t_\mathrm{first}+0.5~\mathrm{Myr}$ to 
increase the number of samples.

\subsection{The Virial Parameter}\label{sec:stat_clumps}

From the virial theorem without the surface terms (Equation (\ref{virialE})), 
a stability criterion can be derived by $\ddot{I}<0$, which is rewritten as
\begin{equation}
    \alpha_\mathrm{tot} = \frac{2\left( \Eth + \Ekin \right) + \Emag}{-\Egrav} < 1.
    \label{alphatot}
\end{equation}
The virial parameter $\alpha_\mathrm{tot}$ is related to that often used in 
observations as a diagnose of the stability of dense clumps and cores \citep{BertoldiMcKee1992},
but it contains the effect of support due to magnetic fields.

Before showing the dependence of $\alpha_\mathrm{tot}$ on $\nth$, $\Mcl$, $V_0$, and $\theta$,
we investigate the velocity structure of the dense clumps that are related to $\Ekin$ 
in Sections \ref{sec:vave} and \ref{sec:ducl}.

\begin{figure}[htpb]
  \centering
 \includegraphics[width=9cm]{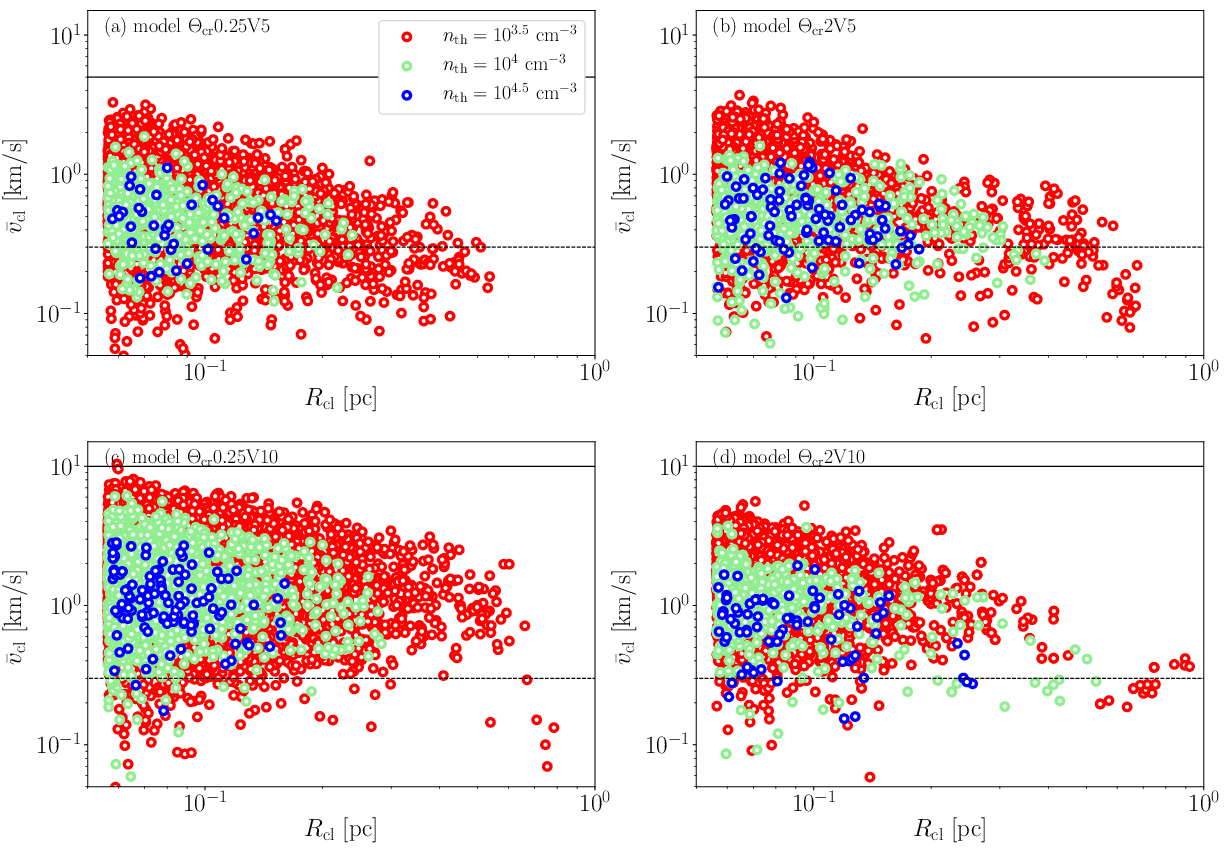}
\caption{
Bulk speeds as a function of the clump size $\Rcl$
at three different $\nth$, (red) $10^{3.5}$, (green) $10^4$, and (blue) $10^{4.5}~\pcc$, 
for models (a)$\mathrm{\Thetacr0.25V5}$, (b)$\mathrm{\Thetacr2V5}$,
(c)$\mathrm{\Thetacr0.25V10}$, and (d)$\mathrm{\Thetacr2V10}$.
The horizontal solid and dashed lines correspond to the collision speed and sound speed in the dense regions, respectively.
}
  \label{fig:clump_vave}
\end{figure}
%
\subsubsection{Bulk Speeds}\label{sec:vave}

In Figure \ref{fig:eneevo}, we found that 
the velocity dispersions of the post-shock layers and their anisotropy in the post-shock layer depend not on $V_0$ significantly but on $\theta/\thetacr$. 
How are these velocity structures imprinted the motion and internal turbulence of the dense clumps?

Figures \ref{fig:clump_vave}
show the bulk speeds of the clumps as a function of the clump size at three different threshold densities 
($n_\mathrm{th}=10^{3.5}$, $10^4$, and $10^{4.5}~\pcc$).
Figures \ref{fig:clump_vave}a and \ref{fig:clump_vave}b show that 
for all the models most clumps have supersonic bulk speeds because the sound 
speed is as low as $0.2-0.3~\kms$ \citep{KI2002}.  
The bulk speeds decrease with the clump size on average although there are large dispersions.

In order to investigate how the bulk velocities of the dense clumps depend on $\theta$ in more detail,
we measure the one-dimensional translational velocity dispersions of the bulk motion of the dense clumps with $\Rcl\le0.2~\mathrm{pc}$ 
that are defined as 
\begin{equation}
    \Delta v_\mathrm{cl} = \sqrt{ \langle \left( \mathbf{v}_\mathrm{cl} 
    - \langle \mathbf{v}_\mathrm{cl}\rangle \right)^2\rangle/3 },
\end{equation}
where the average is taken over all the dense clumps identified with each threshold density.

\begin{figure}[htpb]
  \centering
 \includegraphics[width=9cm]{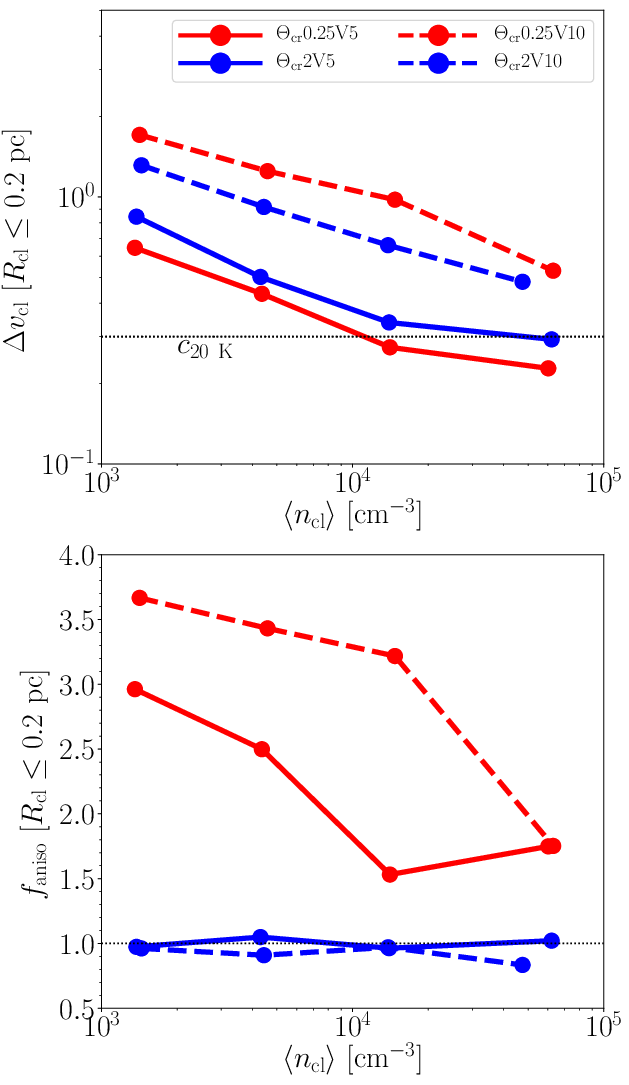}
\caption{
(a) Velocity dispersions of the bulk motion of the dense clumps with $\Rcl\le 0.2~\mathrm{pc}$
for models (solid red)$\Thetacr0.25$V5, (solid blue)$\Thetacr2$V5, (dashed red)$\Thetacr0.25$V10, and (dashed blue)$\Thetacr2$V10
as a function of the mean clump density at four different threshold densities, 
$\nth=10^3,$ $10^{3.5}$, $10^4$, and $10^{4.5}~\pcc$.
In each panel, the horizontal solid and dashed lines correspond to the collision speed and sound speed in the dense regions.
}
\label{fig:clump_vave_pick}
\end{figure}

Figure \ref{fig:clump_vave_pick}a shows $\Delta v_\mathrm{cl}$ 
as a function of the mean clump density $\langle \ncl\rangle$.
All the model shows that $\Delta v_\mathrm{cl}$ decreases at a similar rate as $\langle \ncl\rangle$ increases.
The difference in $\theta$ does not have significant effect on $\Delta v_\mathrm{cl}$.
By contrast, $\Delta v_\mathrm{cl}$ increases roughly in proportion to $V_0$. 
This comes from that the global velocity dispersions are in proportion to $V_0$, but are roughly independent of $\theta$ 
(Figures \ref{fig:eneevo}c and \ref{fig:eneevo}d).  
At $t=t_\mathrm{first}$, the global velocity dispersions are $\sim 1.7~\kms$ for both the models with $V_0=5~\kms$, 
are $\sim 3.4~\kms$ for model $\Thetacr0.25$V10 and $\sim 2.4~\kms$ for model $\Thetacr2$V10.



We further examine how the anisotropy of the velocity dispersions of the post-shock layers shown in Figure 
\ref{fig:eneevo}b carries over the anisotropy of ${\bf v}_\mathrm{cl}$.
To characterize it, $f_\mathrm{aniso}\equiv \Delta v_{x,\mathrm{cl}}/
\Delta v_{\perp,\mathrm{cl}}$
is defined as a representative value of the degree of anisotropy, where 
$\Delta v_{\perp,\mathrm{cl}}^2= (\Delta v_{y,\mathrm{cl}}^2
+ \Delta v_{z,\mathrm{cl}}^2)/2$.

Figure \ref{fig:clump_vave_pick}b shows that 
unlike $\Delta v_\mathrm{cl}$,
$f_\mathrm{aniso}$ depends more strongly on $\theta$ than on $V_0$. 
For both the models with $\theta=2\thetacr$,
$f_\mathrm{aniso}$ is almost constant with $\ncl$ and roughly equal to unity.
This is consistent with Figure \ref{fig:eneevo}b.
By contrast, the anisotropy in the models with $\theta=0.25\thetacr$ are 
more significant toward lower $\ncl$. 
At $\nth=10^3~\pcc$, $f_\mathrm{iso}$ is as high as $\sim 2.8$, which is consistent with 
$\delta v_x/\sqrt{(\delta v_y^2 + \delta v_z^2)/2} \sim 2.8$ for $V_0=5~\kms$ and 
$\sim 3.8$ for $V_0=10~\kms$ at $t=t_\mathrm{first}$ (Figure \ref{fig:eneevo}b).

For $\theta=0.25\thetacr$, 
the larger the collision speed, the higher $f_\mathrm{aniso}$ 
remains until the gas density becomes higher.
This indicates that the ballistic motion of the dense clumps is maintained 
without much deceleration up to higher densities for larger $V_0$ (Figure \ref{fig:clump_vave_pick}).

\subsubsection{Internal Velocity Dispersions}\label{sec:ducl}

\begin{figure}[htpb]
  \centering
 \includegraphics[width=9cm]{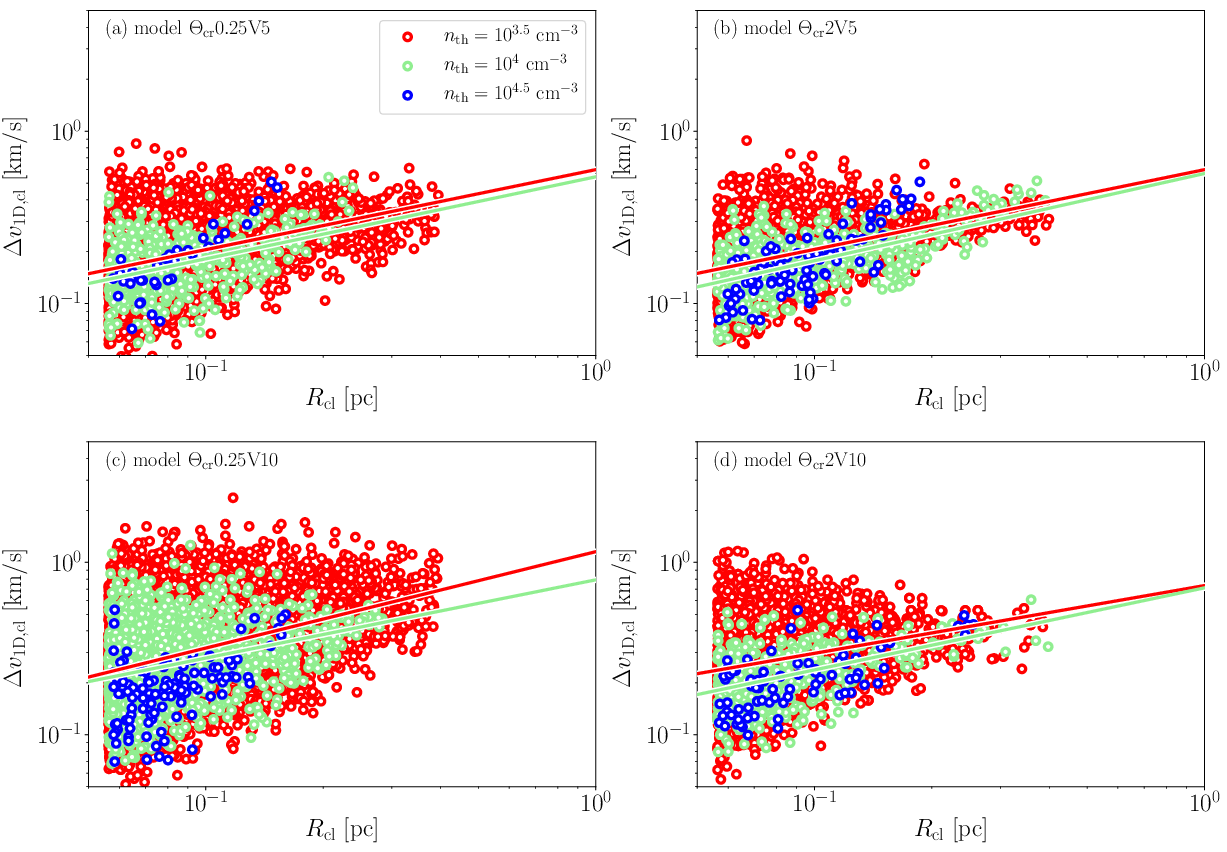}
\caption{
Internal velocity dispersions as a function of the clump size $\Rcl$
at three different $\nth$, (red) $10^{3.5}$, (green) $10^4$, and (blue) $10^{4.5}~\pcc$, 
for models (a)$\Thetacr0.25$V5, (b)$\Thetacr2$V5, (c)$\Thetacr0.25$V10, and (d)$\Thetacr2$V10.
In each panel, the two lines correspond to the lines of the best fit for 
(red) $10^{3.5}$, and (green) $10^{4}~\pcc$.
}
  \label{fig:clump_vdisp}
\end{figure}

We here investigate the size-dependence of the internal velocity dispersion of each dense clump.
The one-dimensional internal velocity dispersions of the clumps $\ducl$ 
are plotted in Figures \ref{fig:clump_vdisp}. 
They increase with the clump size in a power-law manner although the range of the clump sizes is small.
We fit them using a power-law function of 
\begin{equation}
    \ducl = \delta v_0 \left( \frac{\Rcl}{1~\mathrm{pc}} \right)^{a}.
    \label{fit}
\end{equation}
The best fit parameters are listed in Table \ref{tab:fit}. 
In the fitting, the clumps with $\nth=10^{4.5}~\pcc$ are removed 
because a increase of $\ducl$ due to self-gravity is found for $\Rcl>0.1~$pc in Figure \ref{fig:clump_vdisp}.

The power-law indices shown in Table \ref{tab:fit} are consistent with 
$0.4-0.5$ for all the models, which is close to those of shock-dominated turbulence 
\citep[$a=0.5$,][]{Elsasser1976} than incompressible Kolmogrov turbulence ($a=1/3$).
They are consistent with observed values \citep[e.g.,][]{Larson1981,Solomon1987}.

\begin{table*}
    \begin{center}
    \begin{tabular}{|l|l|l|l|l|}
        \hline 
        $\nth~[\mathrm{cm}^{-3}]$ & $\Thetacr0.25$V5 & $\Thetacr0.25$V10 & $\Thetacr2$V5 & $\Thetacr2$V10 \\
        \hline 
        \hline 
        $10^{3.5}$ & 
        $ (0.60\pm0.02)R_\mathrm{1pc}^{0.46\pm 0.01}$ & 
        $ (1.15 \pm 0.03) R_\mathrm{1pc}^{0.56\pm 0.01}$ &
        $ (0.57 \pm 0.04) R_\mathrm{1pc}^{0.51\pm 0.03}$ &
        $ (0.73 \pm 0.04) R_\mathrm{1pc}^{0.39\pm 0.02}$ 
        \\
        \hline
        $10^4$ & 
        $(0.54 \pm 0.06)R_\mathrm{1pc}^{0.48\pm 0.05}$  & 
        $(0.79 \pm 0.07) R_\mathrm{1pc}^{0.45\pm 0.03}$ &
        $(0.57 \pm 0.04) R_\mathrm{1pc}^{0.51\pm 0.03}$ &
        $(0.71 \pm 0.07) R_\mathrm{1pc}^{0.48\pm 0.04}$ 
        \\
        \hline 
    \end{tabular}
    \caption{List of the best-fit formula of the internal velocity dispersions. 
    From the second to fifth columns, 
    the numbers in the parentheses represent $\delta v_0$ in $\kms$.}
    \label{tab:fit}
    \end{center}
\end{table*}

\begin{figure}[htpb]
  \centering
 \includegraphics[width=9cm]{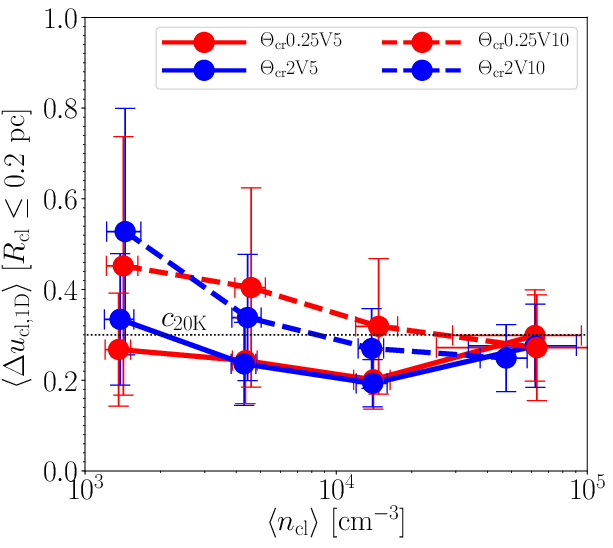}
\caption{
Mass-weighted average of the internal velocity dispersions as a function of the mean clump density 
at $\nth=10^3$, $10^{3.5}$, $10^4$, and $10^{4.5}~\pcc$
for models (solid red)$\Thetacr0.25$V5, (solid blue)$\Thetacr2$V5, 
(dashed red)$\Thetacr0.25$V10, and (dashed blue)$\Thetacr$2V10.
The average is performed over the dense clumps with $\Rcl<0.2$~pc. 
The horizontal and vertical error bars indicate the dispersions of the mean density and internal velocity dispersion 
at each threshold density, respectively.
}
  \label{fig:clump_vdisp_pickup}
\end{figure}

In order to investigate the parameter dependence of $\ducl$ in more detail, 
we plot  $\ducl$ mass-weighted averaged over the clumps with $\Rcl\le 0.2~$pc 
as a function of $\langle \ncl\rangle$ in Figure \ref{fig:clump_vdisp_pickup}.
Like the bulk velocity dispersions $\langle \Delta v_\mathrm{cl}\rangle$, 
$\langle \ducl\rangle$ depends more on $V_0$ than $\theta$.
For $V_0=5~\kms$, the internal velocity dispersions are as low as $\sim 0.2-0.3~\kms$, regardless of 
$\ncl$ for both $\theta=0.25\thetacr$ and $2\thetacr$.
The internal velocity dispersions at $\nth=10^3~\pcc$ increases roughly in proportion to $V_0$.
Their scatters in $\ducl$ also increase with $V_0$.
Unlike for $V_0=5~\kms$, $\ducl$ decreases with $\ncl$ for $V_0=10~\kms$.
As a result, the difference in $\ducl$ between the models with different $V_0$ decreases as $\langle n_\mathrm{cl}\rangle$.
This feature is consistent with the results of \citet{AH2010}.

\subsubsection{Virial Parameters} \label{sec:alpha}

\begin{figure*}[htpb]
  \centering
 \includegraphics[width=18cm]{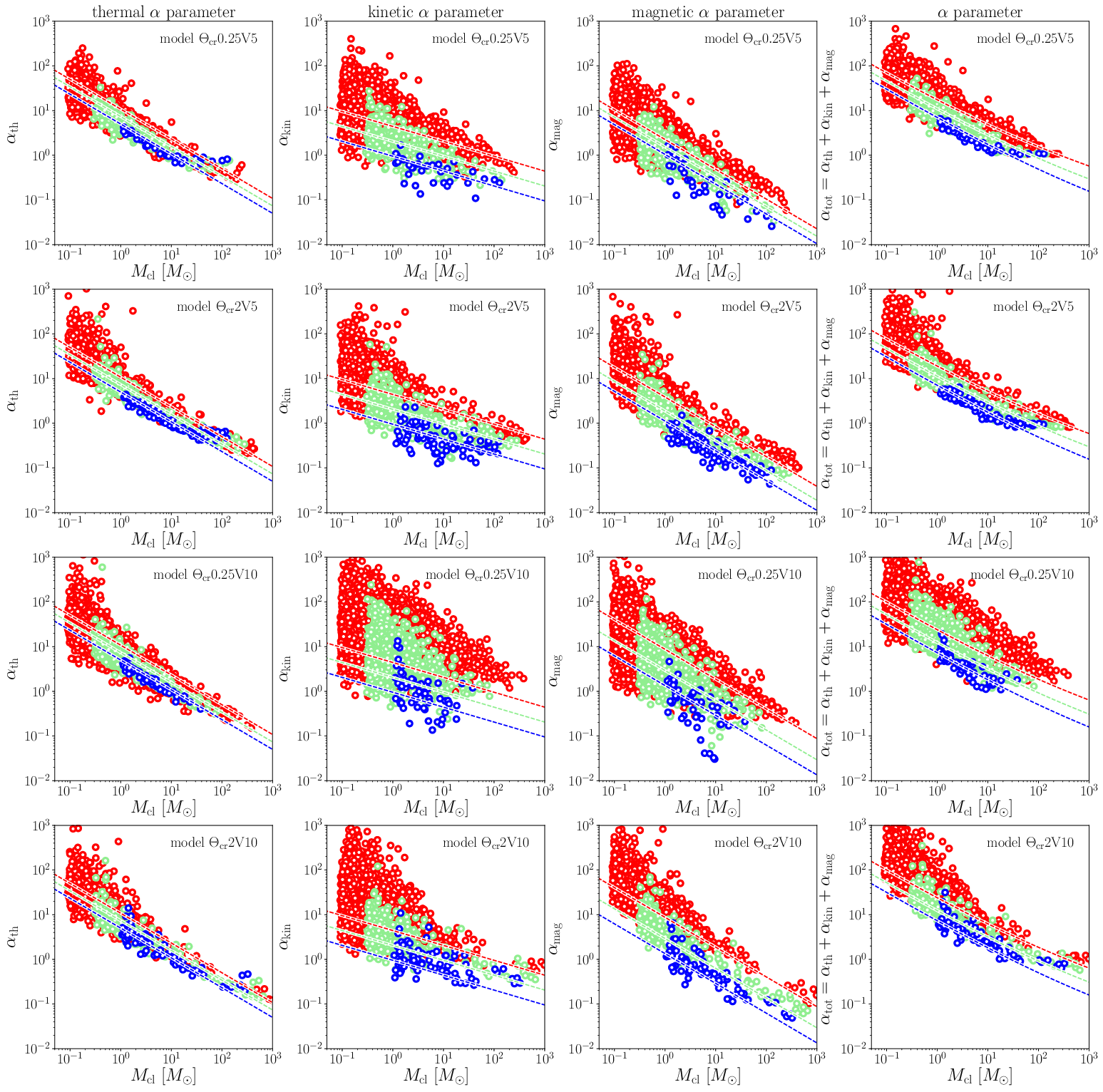}
\caption{
(first column) Thermal, (second column) kinetic, and (third column) magnetic virial parameters as
a function of the clump mass at (red) $\nth=10^{3.5}$, (green) $10^4$, and (blue) $10^{4.5}~\pcc$ for 
models (first row) $\Thetacr0.25$V5, (second row) $\Theta2$V5, (third row) $\Thetacr0.25$V10, and (forth row) $\Thetacr2$V10.
The forth column shows the total virial parameter $\alpha_\mathrm{tot}=\alpha_\mathrm{th}+\alpha_\mathrm{kin}+\alpha_\mathrm{mag}$.
In each panel, the three dashed lines correspond to the predictions from the analytic formula (Equations (\ref{alphath}), 
(\ref{alphakin}), (\ref{alphamag}), and (\ref{alphatotana})).
}
  \label{fig:clump_alpha}
\end{figure*}
The virial parameter (Equation (\ref{alphatot})) is divided into three contributions, thermal, kinetic, and magnetic virial parameters as follows:
\begin{equation}
    \alpha_\mathrm{th} = \frac{2\Eth}{-\Egrav},\;\;
    \alpha_\mathrm{kin} = \frac{2\Ekin}{-\Egrav},\;\;
    \mathrm{and}\;\;
    \alpha_\mathrm{mag} = \frac{\Emag}{-\Egrav}.\;\;
    \label{alphathkinmag}
\end{equation}

Figure \ref{fig:clump_alpha} illustrates the three virial parameters ($\alpha_\mathrm{th}$, $\alpha_\mathrm{kin}$, and $\alpha_\mathrm{mag}$)
as a function of the clump mass at three density thresholds.
One can see that 
the three virial parameters have different $\Mcl$ and $\nth$-dependence.
Focusing on the $\Mcl$-dependence, 
as $\Mcl$ increases, 
$\alpha_\mathrm{th}$ and $\alpha_\mathrm{mag}$ both decrease at a rate of $\propto \Mcl^{-0.6}$ while
$\alpha_\mathrm{kin}$ decreases more slowly following $\propto \Mcl^{-0.3}$.
All the virial parameters decrease with $\nth$, keeping their trend unchanged while
the decreasing rate depends on $\theta$ and $V_0$.


In order to explain the simulation results, 
we here derive analytic formula of $\alpha_\mathrm{th}$, $\alpha_\mathrm{kin}$, and $\alpha_\mathrm{mag}$.
First we consider the thermal virial parameter.
Using Equations (\ref{Ethcl0}) and (\ref{Egrvcl0}),
Equation (\ref{alphathkinmag}) is rewritten as 
\begin{equation}
    \alpha_\mathrm{th,ana} \sim  \frac{5\Rcl}{G\Mcl} \cs^2 = 9c_{0.3}^2 \rho_4^{-1/3} M_1^{-2/3},
    \label{alphath}
\end{equation}
where $c_{0.3}=\cs/0.3~\kms$, $\rho_4 = \rhocl/(10^4 \mu m_\mathrm{H}~\pcc)$, 
and $M_1=\Mcl/1~M_\odot$.
Equation (\ref{alphath}) is essentially the same as that derived by \citet{BertoldiMcKee1992} and \citet{Lada2008}.
The thermal virial parameters are consistent with the predictions from Equations (\ref{alphath})
for all the models (the first column of Figure \ref{fig:clump_alpha}).

The kinetic virial parameter is estimated as
\begin{equation}
    \alpha_\mathrm{kin,ana} \sim \frac{5\Rcl}{G\Mcl} \ducl^2
    = \frac{5\Rcl}{G\Mcl}\delta v_0^2 \left( \frac{\Rcl}{R_0} \right),
    \label{alphakin0}
\end{equation}
where 
the internal  velocity dispersion $\ducl = \delta u_0 \left( \Rcl/R_0 \right)^{0.5}$ is used (Section \ref{sec:ducl}).
Equation (\ref{alphakin0}) becomes 
\begin{equation}
    \alpha_\mathrm{kin,ana} \sim 9 \delta u_1^2 \rho_4^{-2/3} M_1^{-1/3}
    \label{alphakin}
\end{equation}
where we use $\delta u_1=\delta u_0/1~\kms$ 
and $R_0=1~\mathrm{pc}$ for all the models because 
$\ducl$ roughly converges to $\sim 0.6~\kms~(\Rcl/1~\mathrm{pc})^{0.5}$ 
toward higher clump densities, regardless 
of the parameters $V_0$ and $\theta$ (Section \ref{sec:ducl}).
The mass-dependence of $\alpha_\mathrm{kin}$ is weaker than that of $\alpha_\mathrm{th}$, 
indicating that $\alpha_\mathrm{kin}$ dominates over 
$\alpha_\mathrm{th}$ for more massive clumps.
Comparison between $\alpha_\mathrm{kin}$ and the predictions from Equation (\ref{alphakin}) shows that 
Equation (\ref{alphakin}) reproduces the simulation results reasonably well in higher densities although 
deviations are found in lower clump densities for $V_0=10~\kms$ as expected.

The magnetic virial parameter is expressed as 
\begin{equation}
    \alpha_\mathrm{mag,ana} \sim \frac{\Bcl^2}{8\pi \rhocl} \frac{5\Rcl}{3G\Mcl}.
    \label{alphamag0}
\end{equation}
The field strength $\Bcl$ is estimated from 
the $B$-$n$ relation shown in Figure \ref{fig:Bn}.
For the models with $\theta=2\thetacr$, the density-dependence of 
the field strength is characterized by $\ngrav$.
Substituting Equation (\ref{Bana}) into Equation (\ref{alphamag0}) one 
obtains 
\begin{equation}
    \alpha_\mathrm{mag,ana} \sim 
    3 c_{0.3}^{2} \rho_4^{-1/3} M_1^{-2/3} \left( \beta_\mathrm{d}^{-1} + \frac{n_\mathrm{grv}}{\ncl}  \right).
    \label{alphamag}
\end{equation}
This equation shows that the magnetic virial parameter decreases in proportion to $M_1^{-2/3}$. 
This dependence is the same as that of $\alpha_\mathrm{th}$ 
as shown in Figure \ref{fig:clump_alpha}.
The predictions from Equation (\ref{alphamag}) are consistent 
with the simulation results (the third column of Figure \ref{fig:clump_alpha}).

Note that $\alpha_\mathrm{mag}$ is closely related with the mass-to-flux ratio that 
is estimated observationally by the Zeemann measurement 
\citep{Crutcher1999,Heiles2005,Crutcher2010}.
We will discuss an asymptotic property of the mass-to-flux ratio of dense clumps in Section \ref{sec:mass-to-flux}.

Combining Equation (\ref{alphath}), (\ref{alphakin}), and (\ref{alphamag}), one obtains
the analytic expression of the total $\alpha$ parameter,
\begin{eqnarray}
    \alpha_\mathrm{tot,ana} &=& \left( 9 + 3\beta_\mathrm{d}^{-1} + 3\frac{n_\mathrm{grv}}{\ncl} \right) 
    c_{0.3}^2 \rho_4^{-1/3} M_1^{-2/3} \nonumber \\
    &+ & 9 \delta v_1^2 \rho_4^{-2/3} M_1^{-1/3},
    \label{alphatotana}
\end{eqnarray}
where the first term in the right-hand side corresponds to $\alpha_\mathrm{th}+\alpha_\mathrm{mag}$ and 
the remaining term corresponds to $\alpha_\mathrm{kin}$.

It is worth noting the importance of the magnetic support 
in $\alpha_\mathrm{tot,ana}$ (Equation (\ref{alphatotana})).
For clumps with $\rhocl>\rho_\mathrm{grv}$, the field strength 
is roughly determined by the condition $\beta\sim O(1)$.
The third term in the parentheses of the first term 
in the right-hand side of Equation (\ref{alphatotana})
is negligible compared with the other two terms.
In such a dense clump, the magnetic field is only of secondary importance 
in $\alpha_\mathrm{tot,ana}$ because $\alpha_\mathrm{th,ana}$ is three times larger than 
$\alpha_\mathrm{mag,ana}$ when $\beta_\mathrm{d}=1$ 
($\alphath/\alpha_\mathrm{mag} = 2\Eth/\Emag\sim 3\beta$).
This finding suggests that the masses of clumps and cores are determined 
almost independently of the magnetic field strength once $\rhocl$ exceeds $\rho_\mathrm{grv}$.

The diffuse clumps with 
\begin{eqnarray}
    \ncl < \frac{\ngrav}{3}
    &=& 10^3~\pcc \left( \frac{\nave}{10~\pcc} \right)\nonumber \\
    && \left( \frac{V_0}{5~\kms} \right)^2 \left( \frac{\cs}{0.3~\kms} \right)^{-2}
    \label{nclmin}
\end{eqnarray}
are expected to be mainly supported by the shock-amplified magnetic 
fields if the contribution of the kinetic energy is excluded.
This is because the third term dominates over the 
other two terms in the parentheses of 
the first term in the right-hand side of Equation (\ref{alphatotana}).
In our setup, the condition is met when $\ncl<10^3~\pcc$ for $V_0=5~\kms$ and 
$\ncl<4\times 10^{3}~\pcc$ for $V_0=10~\kms$.
This is why magnetically-supported dense clumps with $\nth\ge 10^{4}~\pcc$ are not seen in our results.

\section{Discussions} \label{sec:discussion}

\subsection{Mass-to-Flux Ratio}\label{sec:mass-to-flux}
The mass-to-flux ratio $\mu$ is often used to evaluate the 
importance of magnetic fields in cloud stability.
If the mass-to-flux ratio is larger than the critical value of 
$\mu_\mathrm{cr}=1/2\pi\sqrt{G}$, 
magnetic fields are not strong enough to support clumps against self-gravity
A factor of $1/2\pi$ slightly changes depending on the clump shape
\citep{Strittmatter1966,Mous1976,Nakano1978,Tomisaka1988,Tomisaka2014}.
Observationally $\mu$ is often estimated by taking the ratio between the column density 
and the field strength measured by the Zeeman effect 
\citep{Crutcher1999,Heiles2005,Crutcher2010}.

In terms of $\alpha_\mathrm{mag}$, 
the mass-to-flux ratio is expressed as 
\begin{equation}
    \frac{\mu}{\mu_\mathrm{cr}} \sim \left( \alpha_\mathrm{mag} \right)^{-1/2}
    = \left( \frac{\Emag}{-\Egrav} \right)^{-1/2}.
    \label{mu}
\end{equation}
From Equation (\ref{alphamag}), $\mu/\mu_{\mathrm{cr}}$ is expected to be proportional to 
$\Mcl^{1/3}$, regardless of $\theta$. 
This mass dependence was confirmed in the third column of Figure \ref{fig:clump_alpha}.
In simulations of cloud-cloud collision,
\citet{Sakre2021} found that as the clump mass increases, the magnetic energy increases more slowly than 
the gravitational energy, which is consistent with the positive $M_\mathrm{cl}$ dependence of the mass-to-flux ratio.
Considering compressions along the magnetic field, 
\citet{Banerjee2009} and \citet{Inoue2012}
found a similar mass-dependence of the mass-to-flux ratios which follow $\propto \masscl^{0.4}$. 
\citet{IffrigHennebelle2017} also found a similar relation in semi-global galactic-disk 
simulations, and explained the mass-dependence by a similar argument as in this paper.

\begin{figure}[htpb]
  \centering
 \includegraphics[width=8cm]{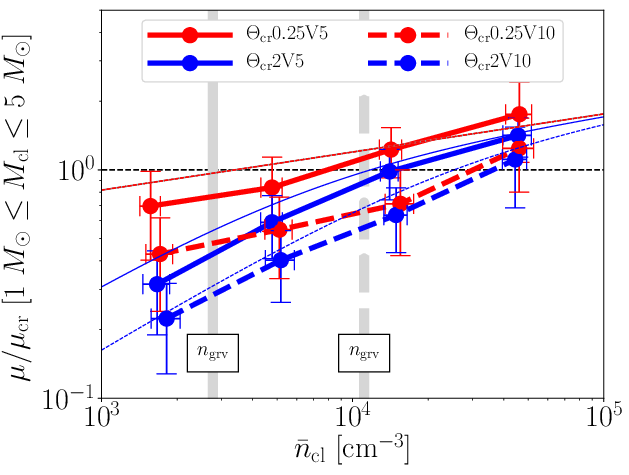}
\caption{
Mass-to-flux ratios averaged over the clumps with $1~M_\odot\le \Mcl\le 5~M_\odot$ as a function of the clump mean density
for models (solid red)$\Thetacr0.25$V5, (solid blue)$\Thetacr2$V5, (dashed red)$\Thetacr0.25$V10, and (dashed blue)$\Thetacr$2V10.
The vertical solid and dashed gray lines indicate $\ngrav$ with $V_0=5~\kms$ and $10~\kms$, respectively.
The thin red line corresponds to the mass-to-flux ratios assuming $\beta=1$.
The thin solid and dashed blue lines indicate the predictions 
from Equation (\ref{alphamag}) with $\Mcl=3~M_\odot$ for $V_0=5~\kms$ and $V_0=10~\kms$, respectively.
}
  \label{fig:mupick}
\end{figure}

We plot the mass-to-flux ratios averaged over $1~M_\odot \le \Mcl \le 5~M_\odot$ 
as a function of the mean clump density in Figure \ref{fig:mupick}.
The $\theta=2\thetacr$ models show that the mass-to-flux ratios are consistent with the predictions from 
the analytic formula (Equation (\ref{alphamag})).
Equation (\ref{alphamag}) predicts 
\begin{equation}
    \frac{\mu_\beta}{\mu_\mathrm{cr}} = 0.6 \beta_\mathrm{d}^{1/2} 
    c_{0.3}^{-1} \rho_4^{1/6} M_1^{1/3}
    \label{mubeta}
\end{equation}
for $\ncl>\ngrav$ and 
\begin{equation}
    \frac{\mu_\mathrm{sh}}{\mu_\mathrm{cr}} = 1.1 \rhoave_1^{-1/2} V_{5}^{-1} \rho_4^{2/3} M_1^{1/3}
    \label{mush}
\end{equation}
for $\ncl<\ngrav$.
The mass-to-flux ratios increase in proportion to $\ncl^{2/3}$ until $\ncl$ reaches $\ngrav$.
When $\ncl$ exceeds $\ngrav$, the mass-to-flux ratios approach the asymptotic line given by the condition $\beta\sim 1$.
The difference in $V_0$ appears in how fast $\mu$ approaches $\mu_{\beta}$.

The difference in $\mu/\mu_\mathrm{cr}$ between the models disappear when $\ncl$ exceeds $\ngrav$ 
because the field strength is determined by the condition $\beta\sim O(1)$ (Figure \ref{fig:Bn}).
Comparison between the simulation results and the predictions from Equation (\ref{alphamag}) with 
$\Mcl=3~M_\odot$  shows that they are consistent for both the models.

For $\ncl >\ngrav$,
the mass-to-flux ratios are insensitive to $\ncl$ (Figure \ref{fig:mupick}) 
since $\mu_\beta\propto \ncl^{1/6}.$
This means that the mass-to-flux ratio does not change significantly once $\ncl$ becomes larger than $\ngrav$.
We thus define the characteristic mass-to-flux ratio $\mu_\mathrm{ch}$ for which $\ncl$ is equal to $\ngrav$,
\begin{equation}
    \frac{\mu_\mathrm{ch}}{\mu_\mathrm{cr}}
    = 0.5 \beta_\mathrm{d}^{1/2}
    c_{0.3}^{-4/3}\rhoave_1^{1/6} V_{5}^{1/3}
    M_1^{1/3}.
    \label{much}
\end{equation}

\begin{figure}[htpb]
  \centering
 \includegraphics[width=8cm]{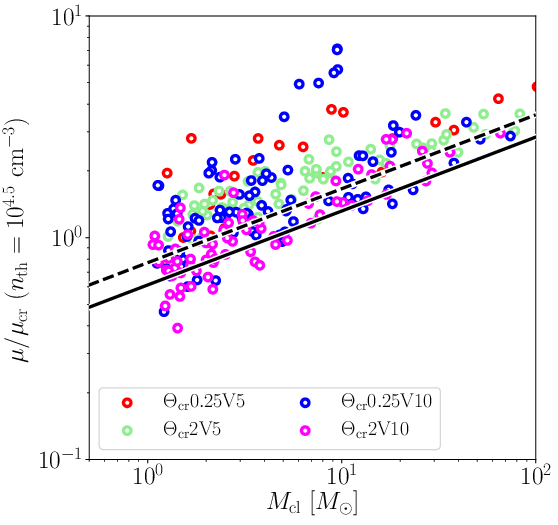}
\caption{
Mass-to-flux ratios of the dense clumps with $\nth=10^{4.5}~\mathrm{cm}^{-3}$
as a function of the clump mass
for models (red)$\Thetacr0.25$V5, (green)$\Thetacr2$V5, (blue)$\Thetacr0.25$V10, and (magenta)$\Thetacr$2V10.
The solid and dashed lines correspond to $\mu_\mathrm{ch}/\mu_\mathrm{cr}$ with $V_0=5~\kms$ and $V_0=10~\kms$, respectively.
}
  \label{fig:mudense}
\end{figure}

Figure \ref{fig:mudense} shows the mass-to-flux ratios of the dense clumps with $\nth=10^{4.5}~\mathrm{cm}^{-3}$,
which are almost independent of the parameters $V_0$ and $\theta$.
This is consistent with what we found in Figure \ref{fig:mupick}.
We compare the mass-to-flux ratios of the dense clumps with $\nth=10^{4.5}~\mathrm{cm}^{-3}$ 
with the predictions from Equation (\ref{much}) in Figure \ref{fig:mupick}.
One can see that they are consistent.
The reason why the measured mass-to-flux ratios are slightly larger than the predictions is 
that the clump density is larger than $\ngrav$.

Interestingly, $\mu_\mathrm{ch}/\mu_\mathrm{cr}$ is roughly equal to unity,
and it is insensitive to the collision parameters $(\nave,V_0)$.
The sound speed is expected to decrease to 0.2~$\kms$ when the gas density increases further, 
$\mu_\mathrm{ch}/\mu_\mathrm{cr}$ increases by a factor of two (Equation (\ref{much})).
We thus expect that 
dense clumps with $\sim 1~M_\odot$ become super-critical 
if their mean densities exceed $\ngrav$ in a wide range of the collision parameters.

\subsection{Internal Alf\'en Mach number} \label{sec:alfmach}

We here show the ratio between the kinetic energy and magnetic energy, or 
Alfv\'en mach number,
\begin{equation}
    {\cal M}_\mathrm{A} = \sqrt{{\Ekin}/{\Emag}}.
\end{equation}

Using Equations (\ref{Bana}) and (\ref{fit}),
one obtains the following analytic formula,
\begin{equation}
    {\cal M}_\mathrm{A,\beta} = 1.2 \delta u_1 c_{0.3}^{-1}
    \rho_4^{-1/6} M_1^{1/6}
    \label{machalfbeta}
\end{equation}
for $n>\ngrav$, and  
\begin{equation}
    {\cal M}_\mathrm{A,sh} = 
    2.3 \delta u_1 V_{5}^{-1} \rhoave_1^{-1/2} \rho_4^{1/3} M_1^{1/6}
    \label{machalfsh}
\end{equation}
for $n<\ngrav$.
If the ram pressure of the colliding flow $\rhoave V_0^2$ is extremely large, 
the {\alfven} Mach number can be small in the low density regions with $n<\ngrav$.
Once the number density exceeds $\ngrav$, Equation (\ref{machalfbeta}) shows that 
the {\alfven} Mach number converges to order of unity. This result is robust because 
it is extremely insensitive to both the clump density and mass.
This feature is consistent with observations by 
\citet{MyersGoodman1988} and \citet{Crutcher1999}.

    \subsection{Structures of Filamentary Clouds}\label{sec:filament}

\begin{figure*}[htpb]
\centering
\includegraphics[width=15cm]{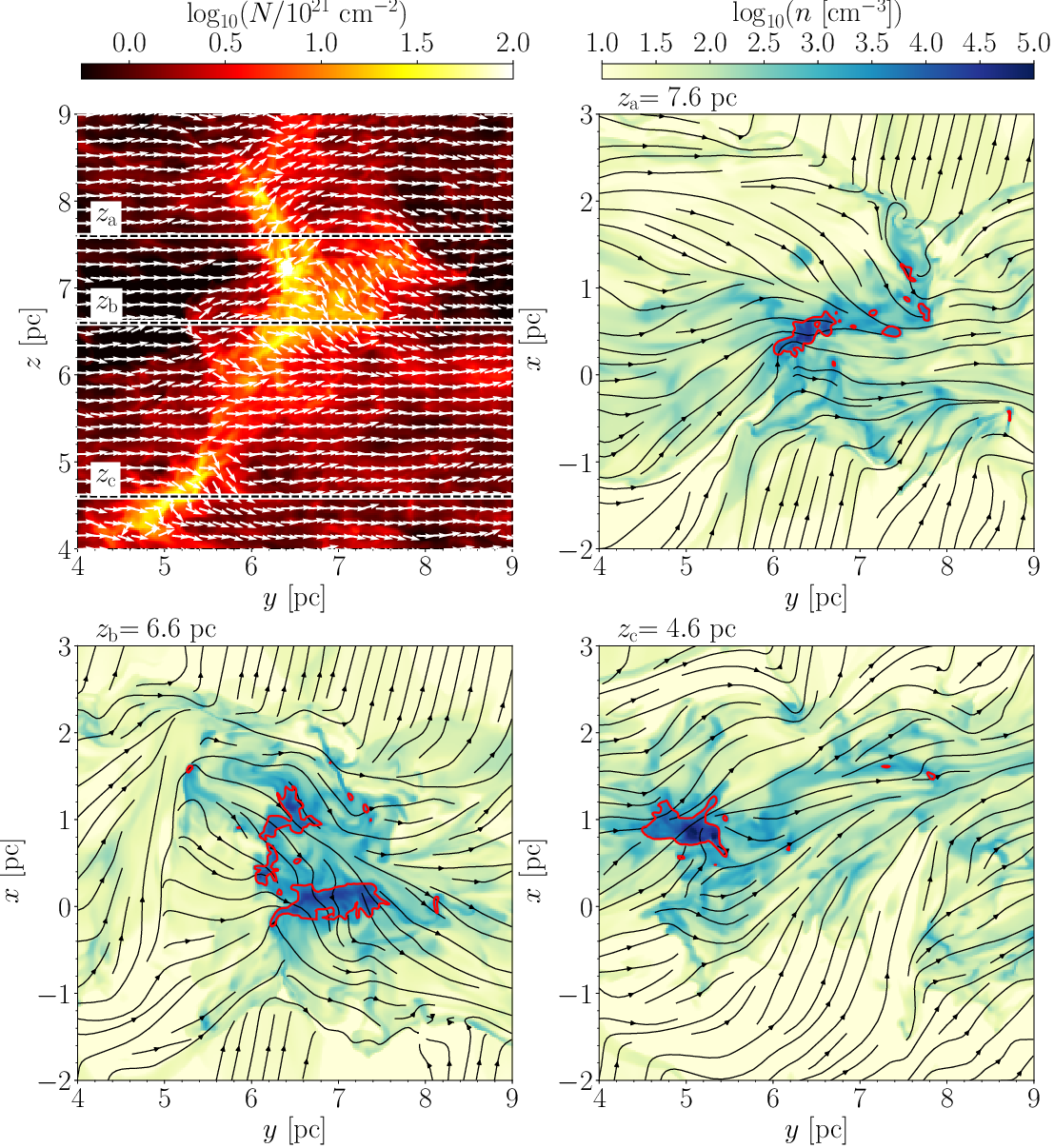}
\caption{
The top-left panel shows the column density map in $4~\mathrm{pc}\le y,z\le 9~\mathrm{pc}$ for model $\Thetacr2\mathrm{V}5$.
The white arrows indicate the directions of the projected magnetic fields density-weighted-averaged along the $x$-axis.
The remaining three panels correspond to the density slices at $z=z_\mathrm{a}=7.6~$pc, $z_\mathrm{b}=6.6$~pc, and $z_\mathrm{c}=4.6$~pc.
The red lines indicate the contours of $n=10^{3.5}~\pcc$.
The black lines show the stream lines of the vector field $(B_y,B_x)$.
}
  \label{fig:filament}
\end{figure*}


In our models, prominent filamentary structures form only for $\theta>\thetacr$ (Figure \ref{fig:colend})
because super-Alfv\'enic turbulence prevents such a coherent structure from forming for $\theta<\thetacr$.
Our results are consistent with the observational results \citep[e.g.,][]{Andre2010} because compressions with $\theta<\thetacr$ 
is extremely rare if compressions are isotropic (Paper I).
In reality, the magnetic fields are not uniform but have different orientations, depending on the location.
Compressions with $\theta<\thetacr$ are unlikely to occur over a wide area.

In this section, the internal structures of the filaments are briefly shown for model $\Thetacr$2V5 although 
they will be investigated in detail in forthcoming papers.
We extract one of the filaments from the simulation results and its column density is shown in the top-left panel of 
Figure \ref{fig:filament}. Overall, the projected magnetic fields are perpendicular to the filament,
which is consistent with observational results \citep[e.g.,][]{Alves2008,Sugitani2010,Chapman2011}.

The three dimensional structure of filaments is not a cylinder but more like a ``ribbon''.
The top-right and bottom panels of Figure \ref{fig:filament} correspond to the density slices at 
$z=z_\mathrm{a}$, $z_\mathrm{b}$, and $z_\mathrm{c}$, whose positions are shown in the top-left panel of Figure \ref{fig:filament}.
The cross sections of the filament are flattened along the local magnetic field direction.
Ribbon-like flattened structures are a natural configuration of strongly-magnetized filaments because the anisotropic nature of the Lorentz force, whose 
direction is perpendicular to the local magnetic field \citep{Tomisaka2014,Auddy2016}.

Interestingly, the elongated direction of the flattened density structures varies significantly
along the filament axis because the directions of the magnetic fields significantly vary 
as shown in the density-slice plots of Figure \ref{fig:filament}.
For instance, at $z=z_\mathrm{b}$, the minor axis of the flattened structure found around $(y,x)\sim (7~\mathrm{pc},0~\mathrm{pc})$ 
is parallel to the line-of-sight direction (the $x$-axis).
That is why the widths of the filament around $z=z_\mathrm{b}$ appear to be larger than those of the other regions 
(the top panel of Figure \ref{fig:filament}).

%

\subsection{A Universal $B$-$n$ Relation}
The $B$-$n$ relation for high densities 
reflects how the magnetic field is amplified during gas contraction.  
The gravitational collapse of a spherical cloud with extremely weak magnetic fields 
leads to the scaling law $B\propto n^{2/3}$ \citep{Mestel1965}.
By contrast, a cloud flattened by the existence of 
the magnetic field undergoes gravitational contraction with $B\propto n^{1/2}$,
indicating that the plasma beta remains constant \citep{Mouschovias1976, Tomisaka1988}.

Our analytic formulae derived in Section \ref{sec:clump} rely on the speculation 
that the field strength is determined by a condition of $\beta \sim O(1)$ in the high density limit,
regardless of $\theta$.


We first present a simple argument of why the central plasma $\beta$ of gravitationally contracting flattened clouds should be $O(1)$.
We then compare the $B$-$n$ relations of our work with those of the previous studies 
in Section \ref{sec:comp_theory}, and those inferred from observational studies in Section \ref{sec:comp_obs}.

\subsubsection{$B$-$n$ Relations of Gravitationally Contracting Flattened Clouds} \label{sec:flatten}

In Section \ref{sec:filament}, we found that the dense clumps have ribbon-like structures flattened along 
the local magnetic field for model $\Thetacr2$V5. 
Also for the other filaments of models $\Thetacr2$V5 and $\Thetacr2$V10, 
similar ribbon-like structures are found.
Thus, at least for $\theta=2\thetacr$, the dense clumps form through gas accumulation along the magnetic fields.

\begin{figure}[htpb]
\centering
\includegraphics[width=8cm]{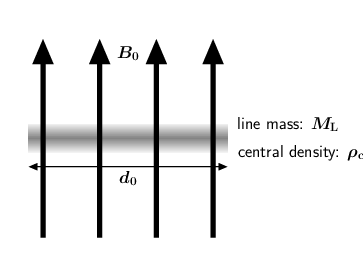}
\caption{
    Schematic picture of the ribbon with the width $d_0$ and line mass $M_\mathrm{L}$.
    The ribbon extends in the direction parallel to the sheet.
    The central density is assumed to be $\rho_\mathrm{c}$, and 
    the density is assumed to be independent of the width.
    The uniform magnetic field (field strength $B_0$) is parallel to the minor axis of the ribbon.
}
  \label{fig:flatten}
\end{figure}

We here investigate how magnetic fields are amplified by self-gravity 
using a simple ribbon model.
As shown in Figure \ref{fig:flatten},
a flattened ribbon-like filament with the central density $\rho_\mathrm{c}$, 
the line mass $M_\mathrm{L}$, and the width $d_0$ is considered.
The uniform magnetic field, whose strength is $B_0$, is perpendicular to the ribbon surface.
The thickness of the ribbon is determined by the thermal Jeans length, 
\begin{equation}
    h = \frac{\cs}{\sqrt{2\pi G \rho_\mathrm{c}}}.
\end{equation}
The relation between $\rho_\mathrm{c}$ and column density $\Sigma = M_\mathrm{L}/d_0$ is 
\begin{equation}
    \rho_\mathrm{c} = \frac{\pi G \Sigma^2}{2\cs^2}.
    \label{rho_sig}
\end{equation}
The flux-freezing condition gives 
\begin{equation}
    B d_0 = \frac{B}{\Sigma} \times M_\mathrm{L} = \mathrm{const.}
    \label{freez}
\end{equation}

We here consider the situation that the ribbon grows by gas accretion along the magnetic field.
Initially, the disk is not massive enough to contract radially due to the magnetic pressure.
In order for the disk to contract radially, the mass-to-flux ratio 
be larger than the critical value,
\begin{equation}
    \frac{\Sigma}{B_0} > \frac{1}{2\pi \sqrt{G}} 
    \label{cond_flat}
\end{equation}
\citep{Nakano1978}.
From this, we define a critical column density of
\begin{equation}
    \Sigma_\mathrm{cr} = \frac{B_0}{2\pi \sqrt{G}}.
\end{equation}
Once the column density of the ribbon reaches this critical value, the ribbon begins to contract radially.
The line mass at this epoch is denoted by $M_\mathrm{L,cr}$,
\begin{equation}
        M_\mathrm{L,cr} = \frac{d_0B_0}{2\pi \sqrt{G}}.
\end{equation}
This is essentially the same as the critical line mass at the large magnetic flux limit derived by \citet{Tomisaka2014}.

Using Equation (\ref{rho_sig}) and (\ref{freez}) ($M_\mathrm{L}\Sigma/B = M_\mathrm{L,cr} \Sigma_\mathrm{cr}/B_0$),
we found that the central plasma $\beta$ evolves as 
\begin{equation}
\beta \sim \frac{8\pi \rho_\mathrm{c}\cs^2}{B^2} = \left( \frac{M_\mathrm{L}}{M_\mathrm{L,cr}} \right)^2.
    \label{beta}
\end{equation}
This result leads to the important conclusion that the plasma $\beta$ of dense clumps 
is order of unity when their gravitational contraction starts.

A similar argument can be done in an oblate cloud flattened along the magnetic field. 
The corresponding expression for the central plasma $\beta$ is given by
\begin{equation}
    \beta \sim \left( \frac{M}{M_\mathrm{cr}} \right)^2,
    \label{betafil}
\end{equation}
where $M$ is the cloud mass and 
\begin{equation}
    M_\mathrm{cr} = \frac{\Sigma \pi R_0^2}{2\pi\sqrt{G}},
\end{equation}
where $R_0$ is the radius of the oblate.

An important conclusion given from 
Equations (\ref{beta}) and (\ref{betafil}) is that the plasma $\beta$ should be larger than unity 
in contracting flattened clouds.

    We should note that the above analytic argument cannot be applied to the models with $\theta=0.25\thetacr$.
    Despite the fact that the dense clumps are not necessarily flattened along the magnetic field, 
    the $B$-$n$ relations of the models with $\theta=0.25\thetacr$ are consistent with the line $\beta \sim O(1)$ (Figure \ref{fig:Bn}).
    More general physics may be hidden.


\subsubsection{$B$-$n$ Relations in Previous Theoretical Studies} \label{sec:comp_theory}

\begin{figure}[htpb]
  \centering
 \includegraphics[width=9cm]{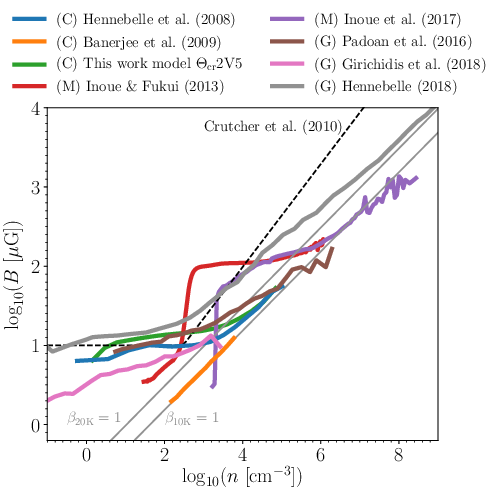}
\caption{
    The mean field strength as a function of the gas density in various studies
    extracted from \citet{Henn2008}, \citet{Banerjee2009}, \citet{Padoan2016}, \citet{Hennebelle2018}, 
    \citet{inoue2013}, \citet{inoue2017}, \citet{Girichidis2018}.
    The $B$-$n$ relations are extracted by using WebPlotDigitizer \citep{Rohatgi2020} 
    except for \citet{inoue2013} and \citet{inoue2017}.
    The two gray lines correspond to $\beta_\mathrm{10K}=1$ and $\beta_\mathrm{20K}=1$.
    The observational prediction in \citet{Crutcher2010}, $B\sim \max(10~\mu\mathrm{G}(n/300~\pcc)^{0.65},10~\mu\mathrm{G})$,
    is shown by the dashed line.
}
  \label{fig:Bn_literature}
\end{figure}

Figure \ref{fig:Bn_literature} shows the $B$-$n$ relations of the simulation studies 
with various setups. 
The references prefixed by (C) \citep{Henn2008,Banerjee2009} 
investigate the colliding flows of atomic gases as in our study.
All the $B$-$n$ relations of the colliding-flow simulations follow the lines of $\beta=1$
for high densities.

Next, our $B$-$n$ relations are compared with those of the colliding-flow of MCs
that is a promising mechanism to form massive cores which will evolve into massive stars
\citep{inoue2013,inoue2017}.
Their field strengths approach the lines of $\beta=1$ as the density increases, 
which is consistent with our finding.
For densities higher than $\sim 10^8~\pcc$, the density-dependence of the field strength becomes weaker than 
$B\propto n^{1/2}$. This may be explained if the mass increases during the gas contraction (see Equation (\ref{beta})).

We should note that the magnetic field for lower densities is stronger than ours because 
their ram pressure is an order of magnitude larger than ours.
The strong magnetic field prevents the gas from becoming gravitationally unstable 
before massive cores form, and 
the dense pre-shock gas provides a sufficient amount of mass to form massive cores. 

We also compare our $B$-$n$ relations with those of supernova-regulated ISM 
MHD simulations \citep{Padoan2016,Hennebelle2018}.
\citet{Padoan2016} carry out a simulation with periodic boundary conditions in a simulation box of $(250~\mathrm{pc})^3$. 
Their results are consistent with our results 
although their simulation setup is quite different from ours.
\citet{Hennebelle2018} consider the stratification of the galactic disk,
and their simulations resolve the gas dynamics from the kpc scale to $0.004~$pc by using a zoom-in technique. 
Their results show that the field strength follows $B\propto n^{1/2}$ for higher densities, 
but it is slightly stronger than that predicted from $\beta=1$ even when the gas density 
is as high as $10^6~\pcc$ if the gas temperature is as low as 10~K. 

\citet{Mocz2017} performed isothermal simulations of driven turbulence with different 
initial field strengths, and demonstrated that the outer regions ($\sim 10^4~$au) 
of the cores have $\beta\sim 1$, regardless of the initial field strength.
The regions of $B\propto n^{2/3}$, where gravitationally collapse isotropically, 
emerge on small scales ($r<10^4~$au)
only when the initial Alfv\'en Mach number is less than unity.
In our simulations, the phase of $B\propto n^{2/3}$ cannot be resolved 
because of the lack of resolution.

\citet{Li2015} found that the mean field strength of clumps does not follow $n^{1/2}$ but 
it is proportional to $n^{0.7}$, which is comparable to those found in observations of the Zeeman measurement
(see Section \ref{sec:comp_obs}).
The mean field strength of their results is $B\sim 20~\mu\mathrm{G}\left( n/10^4~\pcc \right)^{0.7}$, regardless 
of the initial field strength. It is higher than those obtained in our simulations.
An possible reason is the difference of the simulation setups.
In our simulations, dense clumps evolve from the stable stage to the unstable stage through coalescence of 
smaller clumps and the gas accumulation of the atomic gas.
By contrast, \citet{Li2015} prepared a gravitationally unstable gas as the initial condition.
Rapid super-Alfv\'enic contraction of the dense clumps may lead to a strong density-dependence of the field strength ($B\propto n^{0.7}$).

In summary, the plasma $\beta$ in dense regions is order of unity in previous studies conducting colliding flow simulations 
\citep{Henn2008,Banerjee2009,inoue2013,inoue2017}.
However, in other setups, some simulations show consistent results \citep{Padoan2016,Girichidis2018}, but 
some studies are not consistent with ours \citep{Hennebelle2018,Li2015}. 
Further investigations are needed to understand what makes this difference in the high density limit of the $B$-$n$ relation

\subsubsection{$B$-$n$ Relations in Observational Studies} \label{sec:comp_obs}

Analysing the results of Zeeman surveys of Hi, OH, and CN, 
\citet{Crutcher2010} derived an expected field strength as a function of density 
that is given by 
$B_\mathrm{obs} \sim 10~\mu\mathrm{G} (n/300~\mathrm{cm}^{-3})^{0.65}$ 
for $n>300~\pcc$.
It is about three times stronger than our results at $n=10^3~\pcc$, and 
the difference increases with the density, indicating that 
the plasma $\beta$ of the observed clouds is much lower than unity.
\citet{Crutcher1999} found that 
the plasma $\beta$ is as low as $0.04^{+0.03}_{-0.02}$, which 
appears to be contradict with our results.

A reason of this discrepancy 
may come from the fact that the Zeeman surveys 
have mainly been conducted in massive star forming regions.
Suppose that a spherical cloud with a density of $\rho$ and mass of $M$ 
has a magnetic field with a strength of $B$, 
the mass-to-flux ratio is given by 
\begin{equation}
\frac{\mu}{\mu_\mathrm{cri}} = 2\pi \sqrt{G} \frac{M}{B\pi (3M/4\pi \rho)^{2/3}}.
\label{mu0}
\end{equation}
Solving equation (\ref{mu0}) for $M$ using a relation of $B=\cs\sqrt{8\pi\rho\beta^{-1}}$,
one obtains
\begin{eqnarray}
    M &=& 10^3~\mathrm{M_\odot}~
    \left( \frac{\mu/\mu_\mathrm{cri}}{2} \right)^3
    \left( \frac{\beta}{0.04} \right)^{-3/2}\nonumber \\
    &&
    \left( \frac{\cs}{0.2~\mathrm{km~s^{-1}}} \right)^3
    \left( \frac{n}{10^4~\mathrm{cm}^{-3}} \right)^{-1/2}.
\end{eqnarray}
The equation indicates that in order for a cloud with $\beta \sim 0.04$ to be 
gravitationally unstable, it should be massive.
This is consistent with the fact that most data in high density range 
analyzed in \citet{Crutcher2010} are associated with massive star formation.
\citet{inoue2013} and \citet{inoue2017} demonstrated that 
magnetically-supported dense cores formed by cloud-cloud collisions are 
possible sites of massive star formation (Figure \ref{fig:Bn_literature}).
An averaged plasma $\beta$ of such dense cores is expected to be much lower than unity 
as seen in the Zeeman observations although the plasma beta of the central densest region is 
order of unity (Figure \ref{fig:Bn_literature}).

We focuses on low and intermediate mass star formation.
For model $\Thetacr2$V10, a typical plasma $\beta$ and density 
of the post-shock gas are $\sim 0.025$ and $200~\pcc$, respectively,
if we assume that the pre- and post-shock gas are 
uniform and isothermal with a sound speed of $0.3~\kms$.
Only forty-times density enhancement is enough for the plasma $\beta$ to become unity.
Thus, our results are not contradict with the observations measuring 
the field strength through the Zeeman effect.

\subsection{Relative Orientations between Density Structures and Magnetic Fields}

To characterize the relation between the density and magnetic fields,
\citet{Soler2013} define that the angle 
$\varphi$ between the magnetic field and gradient of the density  
using the following expression:
\begin{equation}
    \varphi = \tan^{-1}\left( \frac{|\mathbf{B}\times \mathbf{\nabla}n| }{
    \mathbf{B}\cdot \mathbf{\nabla}n} \right)
    \label{HRO}
\end{equation}
where $\mathbf{\nabla}n$ is calculated using the Gaussian derivative kernel with 
the standard deviation $\Delta x/\sqrt{2}$ to reduce numerical fluctuations.

The two-dimensional version of Equation (\ref{HRO}), where $n$ is replaced by observed surface densities and the direction of $\mathbf{B}$ is 
estimated from polarization observations, is often used to link the simulation and observation studies
\citep[e.g.,][]{Planck2016}.

\begin{figure}[htpb]
  \centering
 \includegraphics[width=9cm]{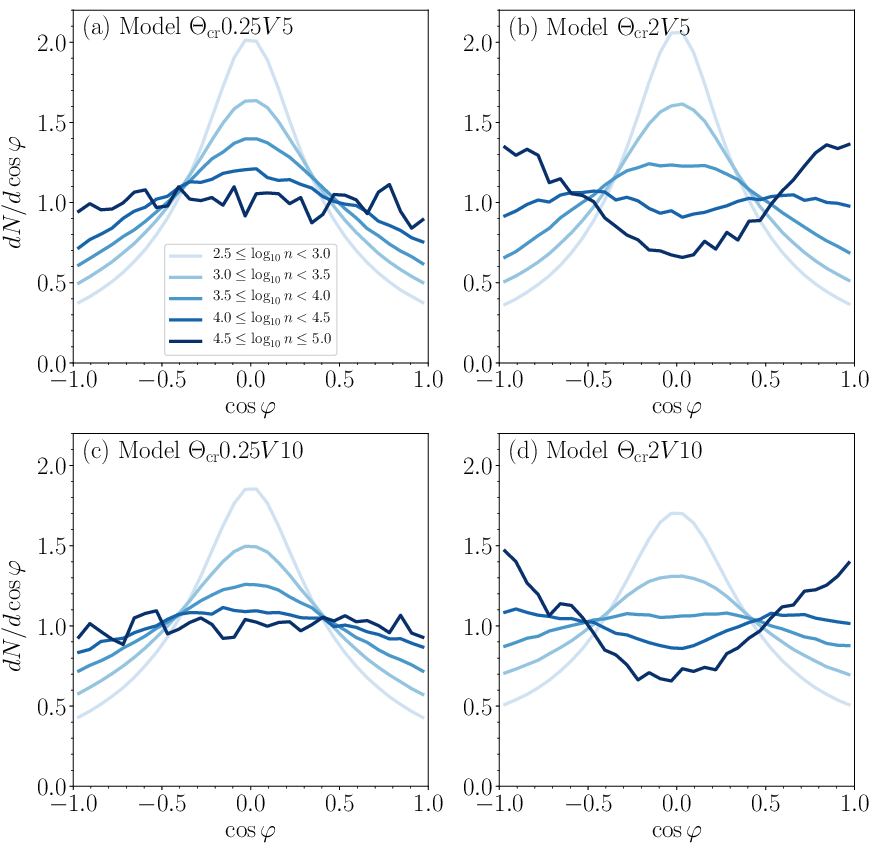}
\caption{
     Three-dimensional HROs obtained from the snapshots at $t=t_\mathrm{first}+0.5~$Myr for models (a) $\Thetacr0.25$V5,
     (b) $\Thetacr2$V5, (c) $\Thetacr0.25$V10, and (d) $\Thetacr2$V10.
     In each panel, the lines correspond to the HROs at the five different density bins.
}
  \label{fig:HRO}
\end{figure}

Figure \ref{fig:HRO} shows 
the histogram of $\cos\varphi$ called Histogram of Relative Orientations (HROs), which was proposed by \citet{Soler2013}.  
For the lowest density bin ($10^{2.5}~\pcc\le n  < 10^{3}~\pcc$),
the HROs have sharp peaks around $\cos\varphi\sim 0$ for all the models, indicating that the density structure tends to be elongated along the magnetic field.
The directions of the density elongation are different between the models with $\theta=0.25\thetacr$ and $2\thetacr$.
For $\theta= 0.25\thetacr$, the strong shear motion, which is generated by the anisotropic super-{\alfvenic} turbulence (Figure \ref{fig:eneevo}c), 
stretches both the density structure and magnetic fields in the collision direction.
By contrast, for $\theta=2\thetacr$, the magnetic fields amplified by the shock compression are perpendicular to the collision direction, and 
the gas is elongated along the magnetic field.

As the density increases,
the peaks around $\cos\varphi \sim 0$ become less prominent while the number of counts around $\cos\varphi\sim \pm 1$ increases for all the models.
The highest density bin ($10^{4.5}~\pcc\le n\le 10^{5}~\pcc$) corresponds to the densest clumps identified in this paper.
Figure \ref{fig:HRO} shows that the HROs at the highest density bin depend on $\theta/\thetacr$.
For $\theta=0.25\thetacr$, 
the HROs become almost flat, indicating that any angles between $\mathbf{\nabla} n$ and $\mathbf{B}$ are equally realized.
For $\theta=2\thetacr$, the magnetic fields tend to be perpendicular to the direction of the elongation of the density structures for $n>10^4~\pcc$. 
This clearly reflects that the filamentary structures are elongated perpendicular to the magnetic field (the right panels of Figure \ref{fig:colend}).

These properties of the HROs are consistent with other simulation studies where 
the flip from the regime of $\mathbf{\nabla} n\perp \mathbf{B}$ to that of $\mathbf{\nabla} n\parallel \mathbf{B}$ 
occurs only in high magnetization cases with $\beta\ll 1$ \citep{Soler2013,Barreto-Mota2021} although  
what determines the critical densities at which the flip occurs is not fully understood theoretically \citep{Pattle2022}. 
For the models with $\theta=2\thetacr$, the critical density ($\sim 10^{4}~\pcc$) roughly agrees with the 
mean density at which the dense clumps with $>1~M_\odot$ change from sub-critical to super-critical (Figure \ref{fig:mupick}).
This is consistent with the results of \citet{Seifried2020}.

\citet{Planck2016,Ade2016} found that 
the relative angle $\varphi$ systematically changes with increasing surface density in a consistent way as the simulation results.
With the clump mass $\sim 10~M_\odot$ around which $\alpha_\mathrm{tot}$ becomes less than unity 
(the bottom panels of Figure \ref{fig:clump_alpha}), the transition density $n\sim 10^{4}~\pcc$ 
corresponds to the surface density 
$N\sim n \times (4\pi/3\times (10~M_\odot)/(1.4 n m_\mathrm{H}~\pcc))^{1/3} \sim 10^{22}~\mathrm{cm}^{-2}$, which appears 
to be consistent with the value $N_\mathrm{H}\sim 10^{21.7}~\mathrm{cm}^{-2}$ obtained from the Planck observations \citep{Ade2016}.
More rigorous comparisons will be made in forthcoming papers by applying synthetic observations to our results.

\subsection{Density Dependence of Velocity Dispersions} 

Figures \ref{fig:clump_vdisp} and \ref{fig:clump_vdisp_pickup} show that the $\ducl$-$R$ relations
depends on the mean clump density although the dynamic range of the clump sizes is less than one decade.
The density dependence is more significant for larger $V_0$.
Observationally, the density-dependence of the $\ducl$-$R$ relation can be investigated 
by using multiple lines of different molecules and isotopes \citep{Goodman1998}.

In the L1551 cloud in Taurus, 
which is an isolated low-mass star-forming region,
\citet{Yoshida2010} do not find any clear deferences in the three line width-size relations 
($\sigma_\mathrm{NT}\sim 0.2~\kms~(L/1~\mathrm{pc})^{0.5}$) 
obtained from the $^{12}$CO ($J=1-0)$, $^{13}$CO ($J=1-0$), and C$^{18}$O ($J=1-0$) lines.  
The obtained velocity dispersions are transonic $\sim 0.2~\kms$ even at $L\sim 1~\mathrm{pc}$, and 
appear to be smaller than a typical line width-size relation \citep[e.g.,][]{Larson1981,Solomon1987}.
This observational result may be qualitatively consistent with the models with $V_0=5~\kms$.

Recently, in the filamentary infrared dark cloud G034.43+00.24, 
\citet{Liu2022} found that the velocity dispersion-size relation in smaller scales ($\sim 0.09~$pc) 
gives systematically lower velocity dispersions than that in larger scales ($\sim 0.6~$pc). 
This may be qualitatively consistent with the models with $V_0=10~\kms$, which show the negative dependence of $\ducl$ on the mean clump density 
(Figure \ref{fig:clump_vdisp_pickup}).

On the other hand, \citet{Liu2022} found that the turbulence spectrum toward a filamentary infrared dark cloud G034.43+00.24 observed in the dense-gas tracer H$^{13}$CO$^+$ 
shows a systematic deviation from the Larson's law and the slope is steeper. 
This may be qualitatively consistent with the models with $V_0=10~\kms$.  
We attribute this deviation from the Larson's law to the gravitational collapse in the dense region.

\subsection{Core Formation} 


We found that for $n>\ngrav$, $\mu/\mu_\mathrm{cr}$ becomes order of unity.
This indicates that the magnetic energy is comparable to the gravitational energy.
Since plasma beta is order of unity for $n>\ngrav$, 
the thermal energy is also comparable to the magnetic energy.
In Section \ref{sec:alfmach}, we found that the Alfv\'en mach number is also order of unity for $n>\ngrav$.
Combining these results suggests that all the energies (thermal, kinetic, magnetic, and gravitational energies)
become equipartition when $n>\ngrav$, regardless of $\theta$.
This property has been found in previous studies.
\cite{LeeHennebelle2019} for instance found that $\Emag$, $\Eth$, and $\Egrav$ reach equipartition in the high-density ends of 
their runs.


\citet{Chen2014} found that core masses do not depend on the upstream magnetic field direction
in their colliding-flow simulations. They estimated the core mass by using a theoretical argument considering gas accumulation along the 
magnetic field. In terms of our notation, 
their estimated core mass corresponds to the 
Jean mass with $n = n_\mathrm{grv} = \nave \left( V_0/\cs \right)^2 \sim 10^5~\pcc$.
From Equation (\ref{much}), the cores with $n\sim \ngrav$ are expected to be super-critical when the core mass is around 1~$M_\odot$. 
Our results suggest that the contribution of the magnetic field to the virial theorem 
is negligible when $n>\ngrav/3=3\times 10^4~\pcc$ (Equation (\ref{nclmin})). 
This is consistent with their results, which show that a typical core mass is independent of magnetic fields.

\citet{Chen2015} investigated the dependence of the core properties on the collision speed and field strength.
The plasma $\beta$ of the cores takes values roughly between 1 and 10, consistent with our results.
They found that the core collapse time is proportional to $V_0^{-1/2}$ \citep[also see][]{Iwasaki2008,Gong2011}.
This appears to be contradict with our results, which show that 
the clump formation time is insensitive to $V_0$ (Table \ref{tab:model} and Figure \ref{fig:mass_nth}).
This discrepancy may come from the difference in the disturbances of the pre-shock gas.
Relatively weak turbulence ($\delta v=0.14~\kms$ in a box of $(1~\mathrm{pc})^3$) was considered in \citet{Chen2015} while
the two-phase structure of the atomic gas is considered in this paper.
In our setting, higher collision speeds do not make the clump formation time shorter because 
stonger post-shock turbulence is driven.

\subsection{Long-term Evolutions} 
In this paper, we focus on the early evolution of the dense clumps since 
our simulations are terminated at 0.5 Myr after the formation of the first unstable clump 
because a star formation treatment is not included and the resolution is not high enough to resolve very dense cores.
In this section, we here discuss whether our results are applicable to the long-term evolution.

In the long-term evolutions, the clumps will grow through gas accretion and denser cores will form.
The analytic formulae of the virial parameters shown in Equations (\ref{alphath}), (\ref{alphakin}), and 
(\ref{alphamag}) are all applicable in the long-term evolutions.
$\alpha_\mathrm{th}$ parameter (Equation (\ref{alphath})) is unlikely to change significantly unless the gas 
is heated up locally by the star formation.
The magnetic energies of the dense clumps are strongly related with the $B$-$n$ relation.
If the magnetic field amplification is caused by the gravitational contraction also in the long-term evolutions, 
Equation (\ref{alphamag}) is also unlikely to change significantly.
It is possible that $\Ekin$ increases as the gravitational potential becomes deeper 
\citep{Arzoumanian2013}, or $\delta u_1$ in Equation (\ref{alphakin}) increases. 
The long-term evolutions with higher resolution are investigated in forthcoming papers.

\section{Summary} \label{sec:summary}

We investigated the formation and evolution of MCs by colliding flows of the atomic gas 
with a mean density of $\nave=10~\pcc$ and field strength of $B_0=3~\mu$G.
We additionally include self-gravity in the simulations done in Paper I and investigate 
the global properties of post-shock layers and the statistical properties of dense clumps.

As parameters, we consider the collision speed and 
the angle between the magnetic field and upstream flow.
As examples of the turbulent-dominated and magnetic field-dominated layers, 
we adopt $\theta=0.25\thetacr$ and $2\thetacr$, respectively.
The critical $\thetacr$ is found in Paper I and defined in Equation (\ref{th_crit}).
We consider two collision speeds of $5~\kms$ and $10~\kms$.

Our findings are as follows:

\begin{enumerate}
     \item The $\theta$-dependence of the physical properties of the post-shock layers affect on the 
           clump formation.
           \begin{itemize}
               \item For $\theta=0.25\thetacr$,  
                   anisotropic super-Alfv\'enic turbulence driven for $\theta=0.25\thetacr$
                   suppresses the formation of self-gravitating clumps.
                   The post-shock layers are significanly disturbed by turbulence, and 
                   no prominent filamentary structures are found.

               \item For $\theta=2\thetacr$,
                   the gas accumulates along the shock amplified magnetic field, and 
                   prominent filamentary structures form.

           \end{itemize}
            Self-gravitating clumps are formed more efficiently for $\theta=2\thetacr$ than 
            for $\theta=0.25\thetacr$.

        \item The difference of $V_0$ does not influence the epoch of the formation of 
            the first unstable clump $t_\mathrm{first}$ significantly although the mass accumulation rate increases in proportion to $V_0$.
            This is because all the energies are in proportional to $V_0^2$, and their ratios are independent of $V_0$.

        \item The parameter dependence of the field strength-density relation appears only 
            for low densities. The mean field strength approaches a univeral relation that determined by $\beta\sim O(1)$
            as the density increases, regardless of values of $\theta$ and $V_0$ (Figure \ref{fig:Bn}).
            The critical density which divides the two asymptotic behaviors 
            is expressed as $\ngrav = \nave (V_0/\cs)^2$ (Equation (\ref{ngrav})).
            We confirm that the $B$-$n$ relations derived in previous studies with different 
            setups approach $\beta\sim O(1)$ in the high-density limit, 
            consistent with our results (Figure \ref{fig:Bn_literature}).

        \item The physical properties of 
            the bulk velocities of dense clumps are inherited from the turbulence structure 
            of the post-shock layers. 
            The models with $\theta=0.25\thetacr$ show that the bulk velocities are 
            biased in the collision direction for less dense clumps.
            As the clump density increases, the anisotropy of the bulk velocities becomes 
            insignificant. 
            Anisotropy in the bulk velocities is not found in the models with $\theta=2\thetacr$.

        \item The internal velocity dispersions of dense clumps roughly obey $\propto \Rcl^{0.5}$, 
            where $\Rcl$ is a typical clump radius, which is defined as $\Rcl= (3\Vcl/(4\pi))^{1/3}$, where 
            $\Vcl$ is the volume of dense clumps.
            The internal velocity dispersions $\ducl$ depend not on $\theta$ but on $V_0$.
            For diffuse clumps with $\sim 10^3~\pcc$, $\ducl$ is roughly in proportion to 
            $V_0$. The larger the collision speed, the faster $\ducl$ decreases with the clump density.
            As a result, the parameter-dependence of $\ducl$ becomes insignifiant for denser clumps.

        \item The virial parameter $\alpha_\mathrm{tot}$ 
            is divided into three contributions from the thermal pressure $\alpha_\mathrm{th}$,
             Raynolds stress $\alpha_\mathrm{kin}$, and 
             Maxwell stress $\alpha_\mathrm{mag}$ in Equation (\ref{alphatot}).
             The thermal and magnetic virial parameters are proportional to $\Mcl^{-2/3}$ and 
            kinetic virial parameter is proportional to $\Mcl^{-1/3}$, where 
            $\Mcl$ is the clump mass.
            We develop an analytic formulae for $\alpha_\mathrm{th}$ in Equation (\ref{alphath}), 
            for $\alpha_\mathrm{kin}$ in Equation (\ref{alphakin}), and for 
            $\alpha_\mathrm{mag}$ in Equation (\ref{alphamag}).
            They explain our results reasonably well.

         \item We found that the mass-to-flux ratios of dense clumps with $\sim 1~M_\odot$ are 
               expected to be order of unity if the clump density is higher than $\ngrav$ in 
               a wide range of the collision parameters which reproduces the observed relations.

\end{enumerate}

\section*{Acknowledgements}
We thank Prof. Eve Ostriker and Masato Kobayashi for fruitful discussions.
We also thank Prof. Tsuyoshi Inoue for providing us their simulation results.
Numerical computations were carried out on Cray XC50 at the CfCA of the 
National Astronomical Observatory of Japan
and supercomputer Fugaku provided by the RIKEN Center for Computational Science (Project ids: hp210164, hp220173.
This work was supported in part by the Ministry of Education, Culture, Sports, Science and
Technology (MEXT), Grants-in-Aid for Scientific Research, 19K03929 (K.I.), 
16H05998 (K.T. and K.I.).
This research was also supported by MEXT as ``Exploratory Challenge
on Post-K computer'' (Elucidation of the Birth of Exoplanets [Second Earth] 
and the Environmental Variations of Planets in the Solar System)
and “Program for Promoting Researches on the Supercomputer Fugaku” (Toward a unified view of the universe: from large scale structures to planets, JPMXP1020200109).

\software{Athena++ \citep{Stone2020}, numpy \citep{Numpy}, 
Matplotlib \citep{Matplotlib}, WebPlotDigitizer \citep{Rohatgi2020}}

\end{document}